\documentclass{jfm}

\usepackage{graphicx}
\usepackage{epstopdf, epsfig}
\usepackage{graphicx}
\usepackage{dcolumn}
\usepackage{bm}
\usepackage{color}
\usepackage{array}
\usepackage{amsmath,amssymb,stmaryrd} 
\usepackage{hyperref}
\usepackage{soul} 
\usepackage{natbib}

\definecolor{red}{rgb}{0.8500, 0.1250, 0.0480} 

\ifCUPmtlplainloaded \else
  \checkfont{eurm10}
  \iffontfound
    \IfFileExists{upmath.sty}
      {\typeout{^^JFound AMS Euler Roman fonts on the system,
                   using the 'upmath' package.^^J}%
       \usepackage{upmath}}
      {\typeout{^^JFound AMS Euler Roman fonts on the system, but you
                   dont seem to have the}%
       \typeout{'upmath' package installed. JFM.cls can take advantage
                 of these fonts,^^Jif you use 'upmath' package.^^J}%
      }
  \else
  \fi
\fi

\ifCUPmtlplainloaded \else
  \checkfont{msam10}
  \iffontfound
    \IfFileExists{amssymb.sty}
      {\typeout{^^JFound AMS Symbol fonts on the system, using the
                'amssymb' package.^^J}%
       \usepackage{amssymb}%
       \let\le=\leqslant  
       \let\ge=\geqslant  
      }{}
  \fi
\fi

\ifCUPmtlplainloaded \else
  \IfFileExists{amsbsy.sty}
    {\typeout{^^JFound the 'amsbsy' package on the system, using it.^^J}%
     \usepackage{amsbsy}}
    {\providecommand\boldsymbol[1]{\mbox{\boldmath $##1$}}}
\fi

\title[Cluster-based feedback control of turbulent post-stall separated flows]
{Cluster-based feedback control of turbulent post-stall separated flows} 

\author{Aditya~G.~Nair\aff{1}
  \corresp{\email{agn13@my.fsu.edu}},
  Chi-An~Yeh\aff{1},~Eurika~Kaiser\aff{2},\\
  Bernd~R.~Noack\aff{3,4,5,6},
  Steven~L.~Brunton\aff{2}
 \and Kunihiko Taira\aff{1}}

\affiliation{\aff{1}Department of Mechanical Engineering, Florida State University, Tallahassee, FL 32310, USA
\aff{2}Department of Mechanical Engineering, University of Washington, Seattle, WA 98195, USA
\aff{3}Laboratoire d'Informatique pour la M\'{e}canique et les Sciences de l'Ing\'{e}nieur, F-91403 Orsay, France
\aff{4}Institut f\"{u}r Str\"{o}mungsmechanik, Technische Universit\"{a}t Braunschweig, D-38108 Braunschweig, Germany
\aff{5}Institut f\"{u}r Str\"{o}mungsmechanik und Technische Akustik, Technische Universit\"{a}t Berlin, D-10623 Berlin, Germany
\aff{6}
Institute for Turbulence-Noise-Vibration Interaction and Control,
Harbin Institute of Technology,
Shenzhen Graduate School, 
University Town, Xili, Shenzhen 518058, 
People's Republic of China}

\pubyear{????}
\volume{???}
\pagerange{???--???}
\begin{document}

\maketitle

 
\begin{abstract}
We propose a novel model-free self-learning cluster-based control strategy for general nonlinear feedback flow control technique, benchmarked for high-fidelity simulations of post-stall separated flows over an airfoil.  The present approach partitions the flow trajectories (force measurements) into clusters, which correspond to characteristic coarse-grained phases in a low-dimensional feature space.  A feedback control law is then sought for each cluster state through iterative evaluation and downhill simplex search to minimize power consumption in flight. Unsupervised clustering of the flow trajectories for in-situ learning and optimization of coarse-grained control laws are implemented in an automated manner as key enablers.  Re-routing the flow trajectories, the optimized control laws shift the cluster populations to the aerodynamically favorable states. Utilizing limited number of sensor measurements for both clustering and optimization, these feedback laws were determined in only $O(10)$ iterations. The objective of the present work is not necessarily to suppress flow separation but to minimize the desired cost function to achieve enhanced aerodynamic performance. The present control approach is applied to the control of two and three-dimensional separated flows over a NACA 0012 airfoil with large-eddy simulations at an angle of attack of $9^\circ$, Reynolds number $Re = 23,000$ and free-stream Mach number $M_\infty = 0.3$. The optimized control laws effectively minimize the flight power consumption enabling the flows to reach a low-drag state.  The present work aims to address the challenges associated with adaptive feedback control design for turbulent separated flows at moderate Reynolds number. 
\end{abstract}



\section{Introduction}
\label{sec:intro}

There is tremendous interest in designing optimal feedback controllers for complex turbulent separated flows to achieve various engineering and technological benefits. Such a feedback control design that autonomously adjusts depending on the state of the flow has advantages in terms of minimizing energy input and robustness to changes in flow conditions \citep{colonius2011control}. Traditionally, excitation of flow instabilities based on open-loop periodic forcing \citep{Greenblatt:PAS00} and feedback control design based on a linear systems approach \citep{Kim:ARFM07} have guided flow control designs. The linear systems framework often relies on linearization of the governing Navier--Stokes equation based on which model-predictive control using adjoint-based optimization techniques or optimal control laws using Riccati-based feedback are designed \citep{sipp2013closed}. However, such control laws derived from linear theory are not able to explore and exploit the nonlinear mechanisms in fluid flows. Also, for real-time fluid flow control, the computational burden is prohibitively large in terms of resources, processing time, and data storage, even for simple geometries.

To alleviate the computational concerns, control strategies are built on low-order dynamical models obtained via model reduction \citep{protas2004linear, pinier2007proportional, barbagallo2009closed, noack2011reduced} or with the use of system identification techniques \citep{huang2008control, bagheri2009input, semeraro2011feedback,illingworth2012feedback,  brunton2016discovering}. Suppressing the large-scale coherent structures in reduced-order models has been shown to mitigate wake unsteadiness, yielding drag reduction in bluff body wake flows \citep{noack2004low, mao2015nonlinear}. Although these methods offer tremendous promise, there are considerable challenges in modeling the interaction of these coherent structures and frequency cross-talk for higher Reynolds numbers, especially in the context of control \citep{luchtenburg2009generalized}. Also, to extract accurate reduced-order models that incorporate nonlinear mechanisms and design control strategies based on them require a high degree of human experience and expertise. 

Alternatively, data-driven flow control holds great potential due to advanced algorithms in machine learning, and modern computational hardware \citep{Brunton2015amr}. Model-free alternatives using genetic programming~\citep{duriez2017machine} have successfully identified control laws in complex cost landscapes in an automated fashion. However, these machine learning control techniques are computationally expensive requiring $\mathcal{O}(1000)$ runs to extract meaningful control laws. Extremum seeking control has shown the ability to adapt to change in flow conditions \citep{ariyur2003real, beaudoin2006bluff} but offers limited flexibility in design of general control laws, optimizing one or few parameters. In the present work, we propose a cluster-based strategy for learning nonlinear feedback control laws directly from coarse-grained fluid flow data to control post-stall separated flow over a canonical airfoil. These control laws have the ability to adapt to nonlinear response from the flow and are deduced in an model-free and automated fashion, allowing for multi-parameter optimization typically with $\mathcal{O}(10)$ runs.
In general, optimization procedures for flow control requires a large number of iterations, but here it scales with the number of discretized clusters, alleviating computational expense for designing feedback control laws for both simulation and experiments. 

The aerodynamic force trajectories are indicators of stall conditions in separated flows. Thus, a low number of force measurements are sufficient to define a feature space without the knowledge of the high-dimensional full flow state. Thus, the fluid flow data of our interest lie in this low-dimensional feature space. Partitioning the feature space into groups sharing similar attributes, called {\it clusters}, the system dynamics can be represented as a linear, probabilistic Markov chain \citep{kaiser2013cluster}. Each cluster corresponds to a characteristic coarse-grained phase of the flow. The Markov transition dynamics between clusters in the feature space translate to the transition between the flow states associated with the clusters. Such a coarse-graining of the feature space into clusters can be leveraged to systematically incorporate nonlinear control mechanisms \citep{kaiser2017cluster}. In the present work, we combine cluster analysis with an iterative optimization procedure to systematically learn global nonlinear feedback control laws. While the control law optimization relies solely on data and is model-free, the resulting dynamics are analyzed using Markov models, providing complementary tools for model-based control design. In this work, Markov models are used only as post-processing analysis to examine the effect of the control strategy on cluster transitions and is not used for feedback control design or control optimization. The control strategy is primarily based on the notion of diverting the force trajectories in the feature space to more favorable regions. In contrast to manipulating the energy transfer between coherent structures to promote patterns with desirable properties, as in model-based approaches, this work aims to re-route the trajectories in an automated fashion to achieve desired flow control objectives. 

\begin{figure}
\begin{center}
\includegraphics[width = 0.99\textwidth]{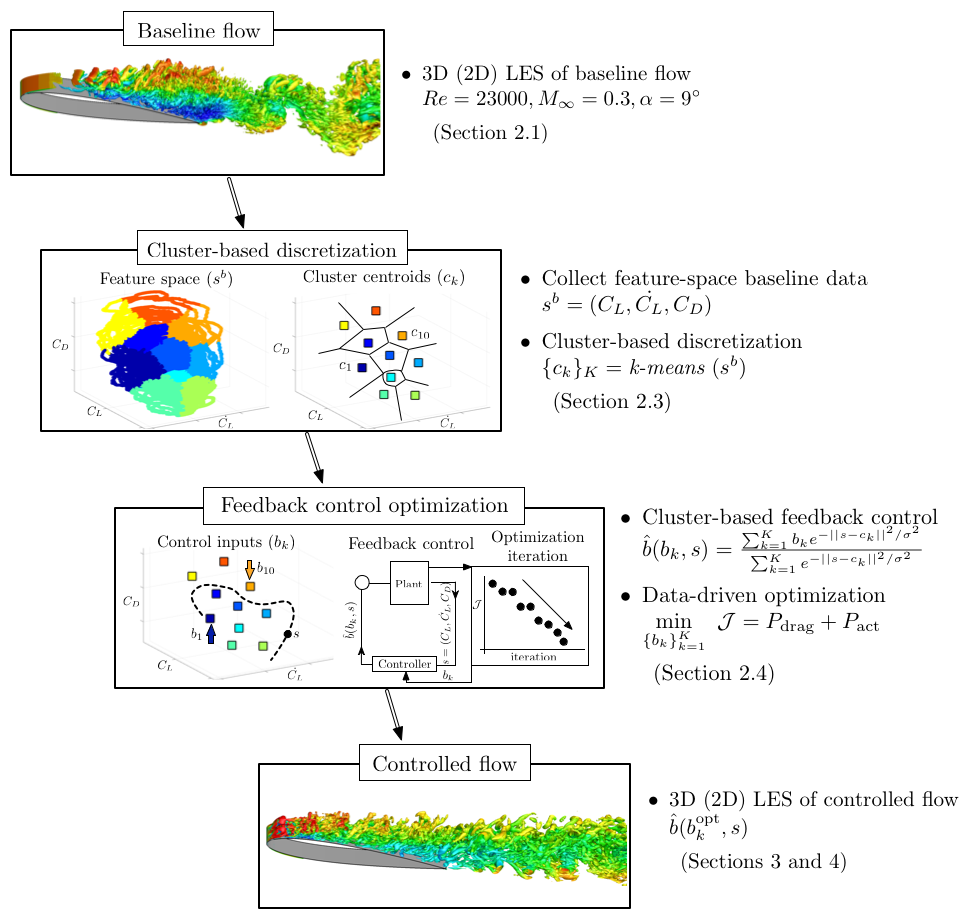} 
\end{center}
\caption{Overview of the presented cluster-based control framework.}
\label{fig:overview}
\end{figure} 

The objective of this work is three-fold; (i) partitioning the baseline flow trajectories (force measurements) into discrete clusters, (ii) optimizing the feedback control law in a model-free manner using the discretized clusters, and finally (iii) analyzing the optimization procedure and found control laws. Unsupervised clustering is used to partition the feature space of trajectories from baseline separated flows into discrete clusters. Assigning a control law to each discretized cluster and actively monitoring and sensing the variables of interest enable the feedback control of flows \citep{kaiser2017cluster}. Our focus is to minimize the power consumption for aerodynamic flight in post-stall flows, i.e., reduce the aerodynamic power with minimum actuation power from data-based principles, thereby improving flight endurance. It must be noted that efforts are not directed towards developing control strategies for full reattachment of the flow but rather towards routing of flow trajectories to minimize power consumption. Control of separated flows is achieved through the optimization of cluster-based control laws to minimize power consumption. Also, the optimization scales with the number of discretized clusters, reducing the computational complexity of the approach, unlike machine learning control techniques. The baseline linear transition dynamics and those to achieve desirable flow behavior with control are examined in post-analysis with networked Markov chain models. Previous efforts have primarily focussed on deducing optimal or suboptimal control laws in physical state-space co-ordinates. Our approach provides a discrete representation of the control law in terms of feature space coordinates, tying the control design directly from data, which improves its applicability to both computational and experimental settings.

We provide an overview of our approach in Figure \ref{fig:overview}. In \S \ref{form:ns}, we discuss the details of the problem setup for baseline simulations. The actuator setup for performing active flow control is described in \S \ref{form:as}. To design feedback control laws for separated flows, baseline feature-space trajectories are collected and discretized into clusters, which is shown in \S \ref{form:cbd}. Each coarse-grained phase of the flow (e.g., each cluster in feature space) is provided with an associated wall-normal blowing/suction jet velocity input for actuation. The path of the controlled trajectories determines the feedback to the flow, enabling the controller to adapt in time. The details of optimization of cluster-based feedback control laws are outlined in \S \ref{form:ofcd}. We demonstrate the effectiveness of the approach for two--dimensional (2D) separated flow over an airfoil in \S \ref{sec:2Dresults}. An extension of the control framework with addition of constraints is then demonstrated for three--dimensional (3D) separated flow over an airfoil in \S \ref{sec:3Dresults}. At last, concluding remarks are offered in \S \ref{sec:cr}.


\section{Cluster-based control framework}
\label{sec:ctrla} 

Once the problem and actuator setup for flow control are determined, a general outline of the cluster-based control framework consists of the following steps, as illustrated in Figure \ref{fig:overview}: (a)~Identify the structure of the control law, (e.g., on-off, multi-frequency forcing, MIMO feedback), (b)~select a feature space consistent with the expected actuation mechanism and online available sensors, (c)~cluster the trajectories in the feature space for unforced benchmark (baseline flow) to parameterize the control law, (d)~choose an exploration-exploitation algorithm (brute force search in case of on-off control; Monte Carlo in case of many expected minima; Simplex search in case of smooth control landscape), (e)~analyze the found control mechanisms and optimization convergence with Markov models and proximity maps, and finally (f)~continue optimizing by adding clusters with respect to the best performing control laws. The specific details of each of these steps are discussed below.

\subsection{Problem setup}
\label{form:ns}

\begin{figure}
\begin{center}
\includegraphics[width = 0.99\textwidth]{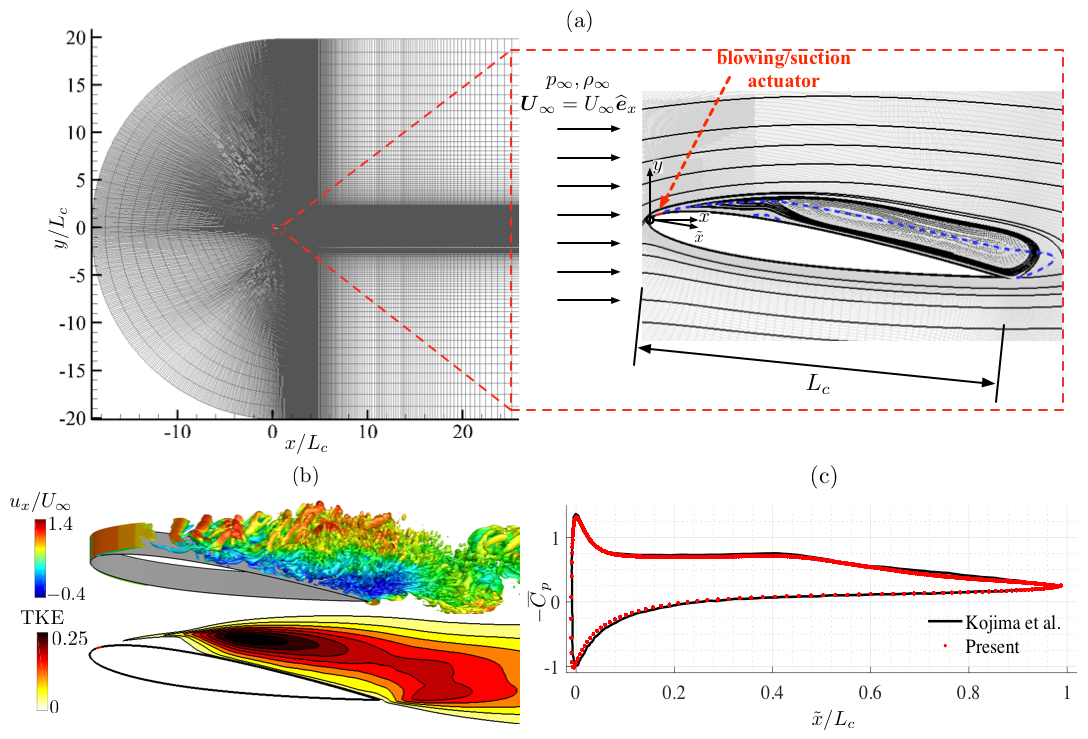} 
\end{center}
\caption{(a) The $x-y$ plane of the computational domain (left) and the near field of a NACA 0012 airfoil at $\alpha = 9^\circ$ (right) with the streamlines for 3D spanwise-periodic baseline flow. The actuator location is indicated in red. The blue dashed line shows the contour line for $\bar{u}_x/U_\infty = 0$. (b) Instantaneous flow field (highlighted by Q-criterion) colored by streamwise velocity and turbulent kinetic energy (TKE). (c) Time-averaged coefficient of pressure distributions on suction and pressure surfaces of the airfoil for 3D baseline flow.}
\label{fig:setup}
\end{figure}

We consider 2D and 3D separated flows over a NACA 0012 airfoil at an angle of attack $\alpha = 9^\circ$, with Reynolds number $Re = U_\infty L_c/\nu = 23,000$ and Mach number $M_\infty = U_\infty/a_\infty = 0.3$. Here, $U_\infty$ is the free stream velocity, $L_c$ is the chord length, $\nu$ is the kinematic viscosity, and $a_\infty$ is the free stream speed of sound. The free stream density is $\rho_\infty$ and free stream pressure is given by $p_\infty$. The streamwise, normal and spanwise coordinate directions are denoted by $x,y$ and $z$, respectively. Henceforth, $u_x, u_y$, and $u_z$ stand for streamwise, normal and spanwise velocity, respectively. To simulate the separated flows, compressible LES is performed. Spanwise periodicity is enforced in the 3D simulations. The computational setup is summarized in Figure~\ref{fig:setup}. The drag coefficient $C_D$, lift coefficient $C_L$, and pressure coefficient $C_p$ are defined as 
\begin{equation}
C_D = \frac{F_x}{\frac{1}{2}\rho_\infty U_\infty^2 A}, ~~~C_L = \frac{F_y}{\frac{1}{2}\rho_\infty U_\infty^2 A}, ~~~C_p = \frac{p-p_\infty }{\frac{1}{2}\rho_\infty U_\infty^2},
\label{forcecoeff}
\end{equation}
where $F_x$ and $F_y$ are the drag and lift forces on the airfoil, $A = L_c w$ is the projected area, $w$ is the width of the airfoil, and $p$ is the pressure at the airfoil surface. The details of the setup, flow visualizations as well as numerical validation, as shown in Figure~\ref{fig:setup}(a), (b) and (c), respectively are discussed in Appendix A.

\subsection{Actuator setup}
\label{form:as}

\begin{figure}
\begin{center}
\includegraphics[width = 0.9\textwidth]{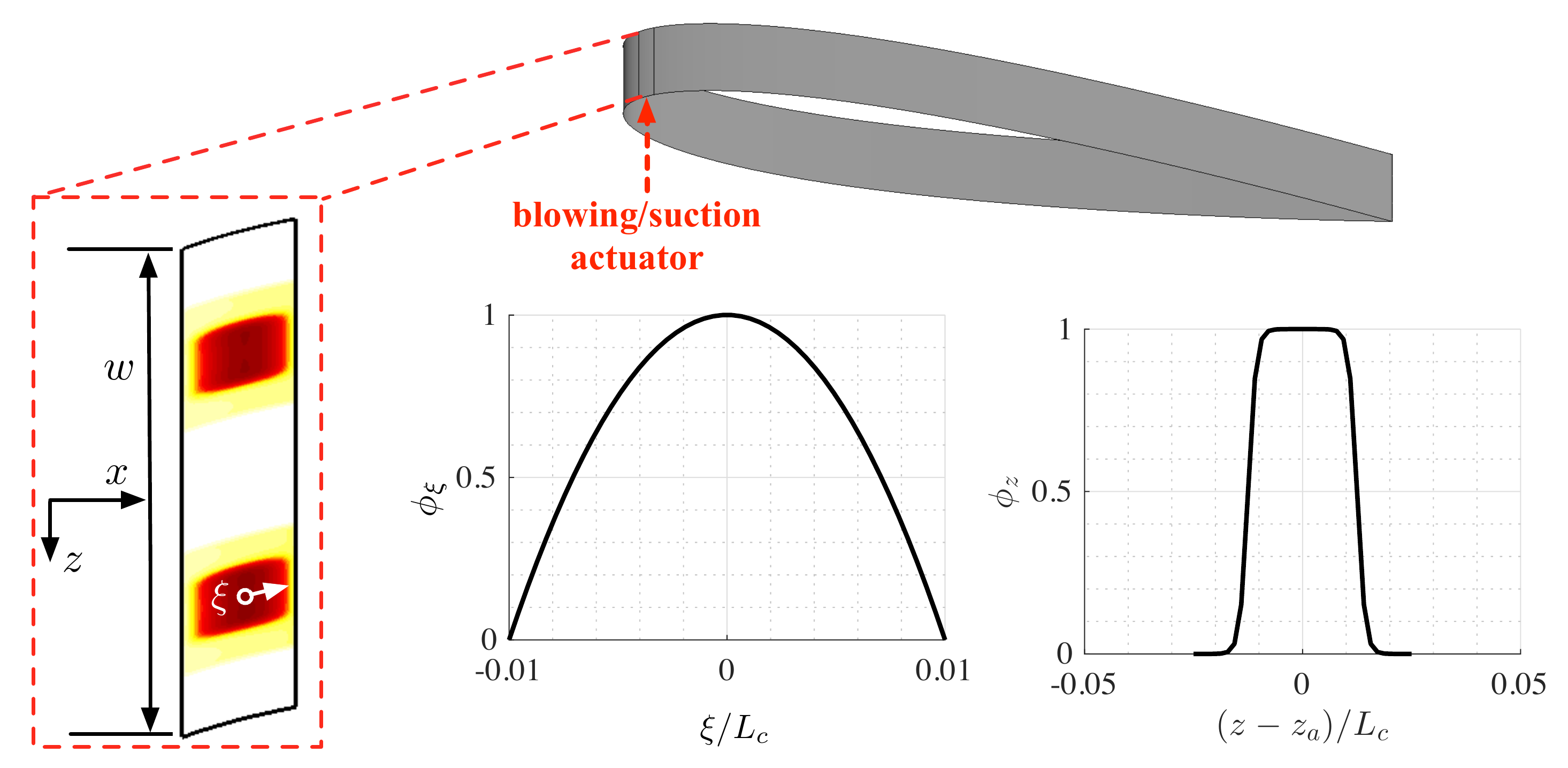} 
\end{center}
\caption{The blowing/suction actuator setup for 3D flow control with velocity profiles in the surface tangential $(\phi_\xi)$ and spanwise $(\phi_z)$ directions.}
\label{fig:setup1}
\end{figure}

To perform flow control, a blowing/suction actuator is centered at $x_a/L_c = 0.03$ in the streamwise direction on the suction side of the airfoil, as shown by red surface in Figure \ref{fig:setup}(a) (right). This location is chosen such that it is upstream of the time-averaged separation point ($x_s/L_c = 0.032$ and $0.037$ respectively for 2D and 3D flows). The actuator setup is further elaborated in Figure \ref{fig:setup1}. Let $\xi$ be the surface tangential direction from the actuator center. The actuator width is $2\xi_a = 0.02 L_c$. A wall-normal velocity component ($u_\text{jet}$) with a parabolic spatial profile $(\phi_\xi)$ is prescribed as an actuator velocity boundary condition to impose blowing/suction. For 3D flow control, two actuator slots are placed in the spanwise direction, each centered at $z_a/L_c = -0.05$ and $0.05$, respectively with a width of $0.025L_c$, similar to the work by \cite{munday2017effects}.  A hyperbolic tangent function $(\phi_z)$ is used for the spanwise jet velocity profile to smoothen the velocity discontinuity at the edge of the slots. The wall-normal velocity is prescribed as 
\begin{equation}
u_\text{jet} = b(s)~\phi_\xi(\xi)\phi_z(z),
\label{ujet}
\end{equation}
where $b$ is the forcing amplitude dependent on the flow state variable $s$ described below. To determine the forcing amplitude, we follow a cluster-based control strategy.

\subsection{Cluster-based discretization}
\label{form:cbd}

Determination of an appropriate feedback control law requires the knowledge of the flow state at each instant in time. For LES at a moderate Reynolds number, the description of a flow typically necessitates millions of degrees of freedom equal to the number of grid points times the number of variables considered. Use of a full-state feedback is thus prohibitive. Instead of utilizing the high-dimensional full state information, we consider a feature space of the baseline flow comprised of a limited number of observables. 

The amplitude and phase dynamics for bluff body flows are captured well by the lift coefficient $C_L$ and its time derivative $\dot{C}_L$, while the drag coefficient $C_D$ captures the mean shift effects \citep{Noack:JFM03, taira2018phase}. These reduced number of observables define a three-dimensional feature set denoted by $s = s(t) = (C_L(t),\dot{C}_L(t),C_D(t))$. A similar feature space was also utilized in the work by \cite{loiseau2017sparse} to deduce sparse, reduced-order, nonlinear models for a two-dimensional flow over a circular cylinder at $Re = 100$. Generally, as the complexity of the flow increases with increasing Reynolds number, the flow becomes higher-dimensional and modeling the interaction terms becomes infeasible to extract a low-dimensional framework. In this work, we pursue a partitioning of the feature space into clusters to identify coarse-grained phases of the flow, which defines a model-free parametrization of the feedback control law, suitable for complex flows. The use of clusters enables to reduce the flow dimension to the order of the attractors. The model-free parametrization can then be constrained to the actual attractor to avoid the curse of dimensionality, which further motivates the data-driven clustering analysis. 
 
The cluster analysis partitions a set of objects or observations with common characteristics into few distinct groups known as {\it clusters}. There are many available clustering techniques including the spectral clustering, hierarchical clustering, and subspace clustering, centroid-based clustering each with their individual pros and cons \citep{rokach2005clustering}. Cluster analysis is commonly used to find a natural grouping among data. Here we employ clustering to discretize or {\it coarse-grain} the feature space of baseline trajectories to deduce feedback control laws. The clustering groups the flow states with similar aerodynamic characteristics, e.g., high-drag states and low-drag states of the flow. 

\begin{figure}
\begin{center}
\includegraphics[width = 0.9\textwidth]{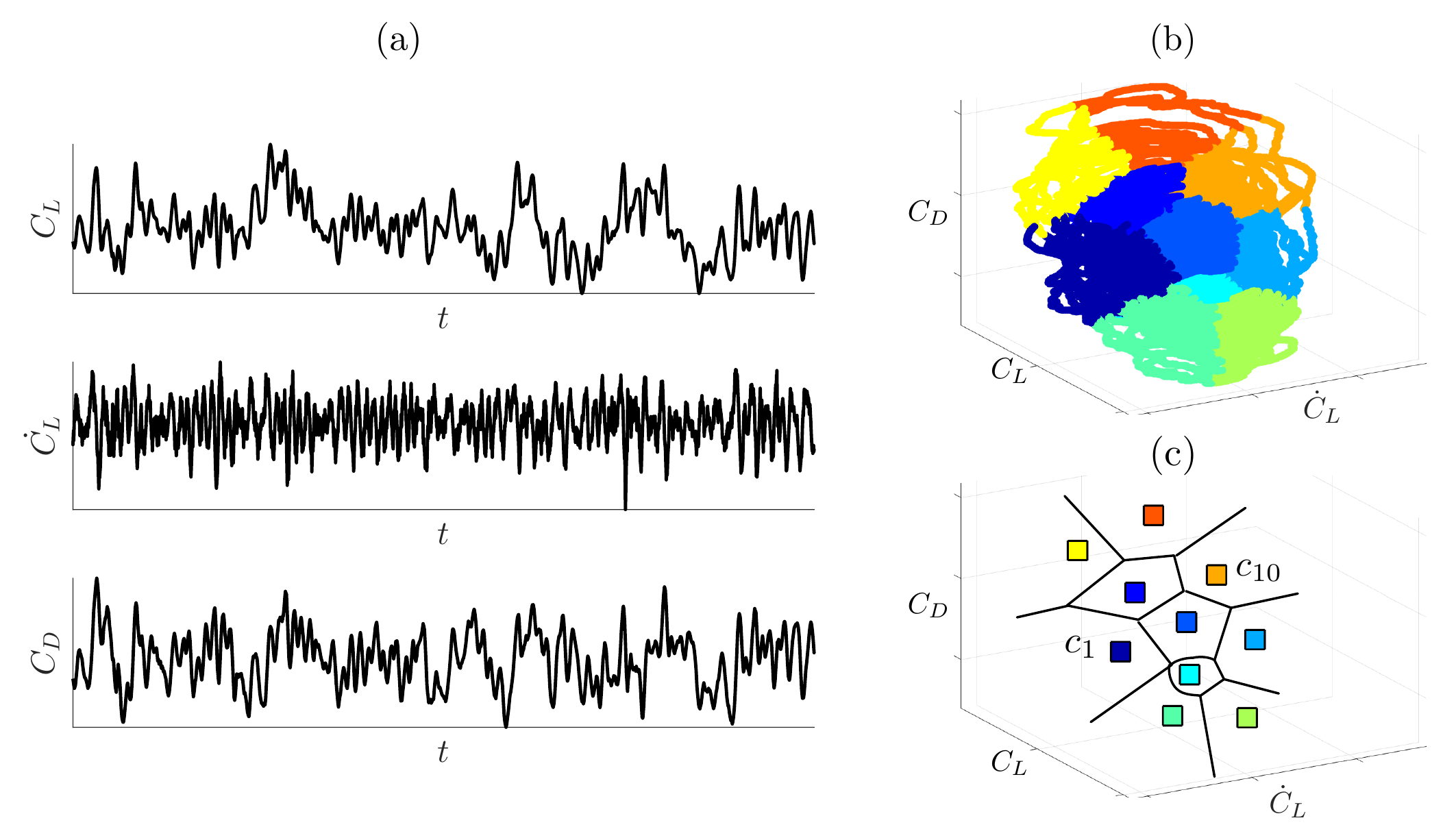} 
\end{center}
\caption{A schematic of the clustering procedure; (a) Time-series of baseline trajectory $s^b(t) = (C_L(t),\dot{C}_L(t),C_D(t))$ collected from baseline LES. (b) Cluster-based discretization of the feature space and (c) corresponding cluster centroids using k-means clustering algorithm.}
\label{fig:setup2}
\end{figure}

One of the most popular centroid-based clustering technique is the {\it k-means} algorithm \citep{lloyd1982least} which is an unsupervised classification algorithm where observations are partitioned into $K$ representative clusters $\{\mathcal{C}_k\}_{k=1}^K$. Each set of observations belonging to a cluster $\mathcal{C}_k$ is represented by its corresponding cluster centroid $c_k$, which is computed as the mean over all observations belonging to this cluster. These cluster centroids represent the mean behavior of the clusters. The clustering analysis is summarized in Figure \ref{fig:setup2}. 

We collect the time-series of baseline flow trajectory, $s^b(t) = (C_L(t),\dot{C}_L(t),C_D(t))$ as shown in \ref{fig:setup2}(a) which forms the measurements to define our feature space. Each feature space co-ordinate is associated with a characteristic baseline velocity field $\boldsymbol{u}^b$. Here, the superscript $b$ denotes the baseline flow. For a set of cluster-based centroids $\{c_k\}_{k=1}^K$, the within-cluster variance ($J_w$) and the inter-cluster variance ($J_i$), as defined in the work by \cite{goutte1999clustering}, are given by
\begin{equation}
J_w = \frac{1}{N} \sum_{k = 1}^K \sum_{s^b \in \mathcal{C}_k} ||s^b - c_k||^2 \quad \text{and} \quad J_i = \frac{1}{N} \sum_{k = 1}^K N_k ||c_k - \bar{c}||^2.
\label{clust_opt1}
\end{equation}
Here, $N$ is the total number of measurements in the baseline trajectory and $N_k$ is the number of measurements present in cluster $\mathcal{C}_k$. The cluster centroid and centroid of the entire trajectory are given by $c_k \equiv \frac{1}{N_k}\sum_{s^b \in \mathcal{C}_k} s^b$ and $\bar{c} = \frac{1}{N}\sum_{k = 1}^K N_k c_k$, respectively. 

The optimal number of clusters can be determined by either minimizing the within-cluster variance or maximizing the inter-cluster variance. Given an ensemble of observations in terms of a baseline flow trajectory $s^b(t)$, the optimal set of cluster-based centroids $\{c_k\}_{k=1}^K$ is obtained by solving an optimization problem that minimizes the within-cluster variance
\begin{equation}
(c_1,...,c_K) = \mathop{\text{arg min}}_{\mathcal{C}}~ J_w.
\label{clust_opt}
\end{equation}
This yields a set of $K$ clusters, $\mathcal{C} = \{\mathcal{C}_1,...,\mathcal{C}_K\}$, each with a centroidal representative state $c_k$.  The cluster-based discretization of the feature space and the corresponding centroids are shown in Figure \ref{fig:setup2}(b) and (c), respectively. In the work of \citet{kaiser2013cluster}, this has led to a cluster-based reduced-order model (CROM) by modeling the transitions as a Markov process \citep{norris1998markov}. There exist an important connection to the linear Liouville equation and the associated Perron-Frobenius operator, which propagate an ensemble of trajectories in phase space. In particular, CROM is founded on discretization of the Perron-Frobenius operator and yields a coarse-grained linear model in the probability space, represented by the Markov transition matrix. Here, we examine the Markov transitions for the baseline and controlled flows as an analysis tool to reveal how the optimized control law modifies transition probabilities, i.e., redirects trajectories in phase space.

A tradeoff between complexity of the cluster-based representation and data compression determines an optimal choice of the number of clusters $K$ \citep{chiang2010intelligent}. The appropriate number of clusters can be determined using an elbow method or the F-test \citep{lomax2013statistical}. The F-test uses the ratio of inter-cluster variance $J_i$ to the total variance $J$, which is typically maximized. We chose $J_i/J\ge0.9$ for the present analysis which yields $K = 10$.  Thus, each dataset of trajectories for the 2D and 3D baseline flows are partitioned into $K = 10$ clusters. With this clustered feature space discretization, we discuss our flow control design below. 

\subsection{Optimized feedback control design}
\label{form:ofcd}

Traditional approaches for flow control involve modeling and manipulating the interaction among a few relevant coherent structures. Models are often constructed by representing the flow with a reduced number of coherent structures or spatial modes to model the dynamical behavior of the flow. However, these modes are generally limited to the coverage of the data they have been extracted from. By applying actuation, the dynamical behavior of the flow often changes and thus leads to a change in the fidelity of the model description. Hence, modeling errors and changes in the flow with control due to nonlinearity should be carefully considered. 

\begin{figure}
\begin{center}
\includegraphics[width = 0.725\textwidth]{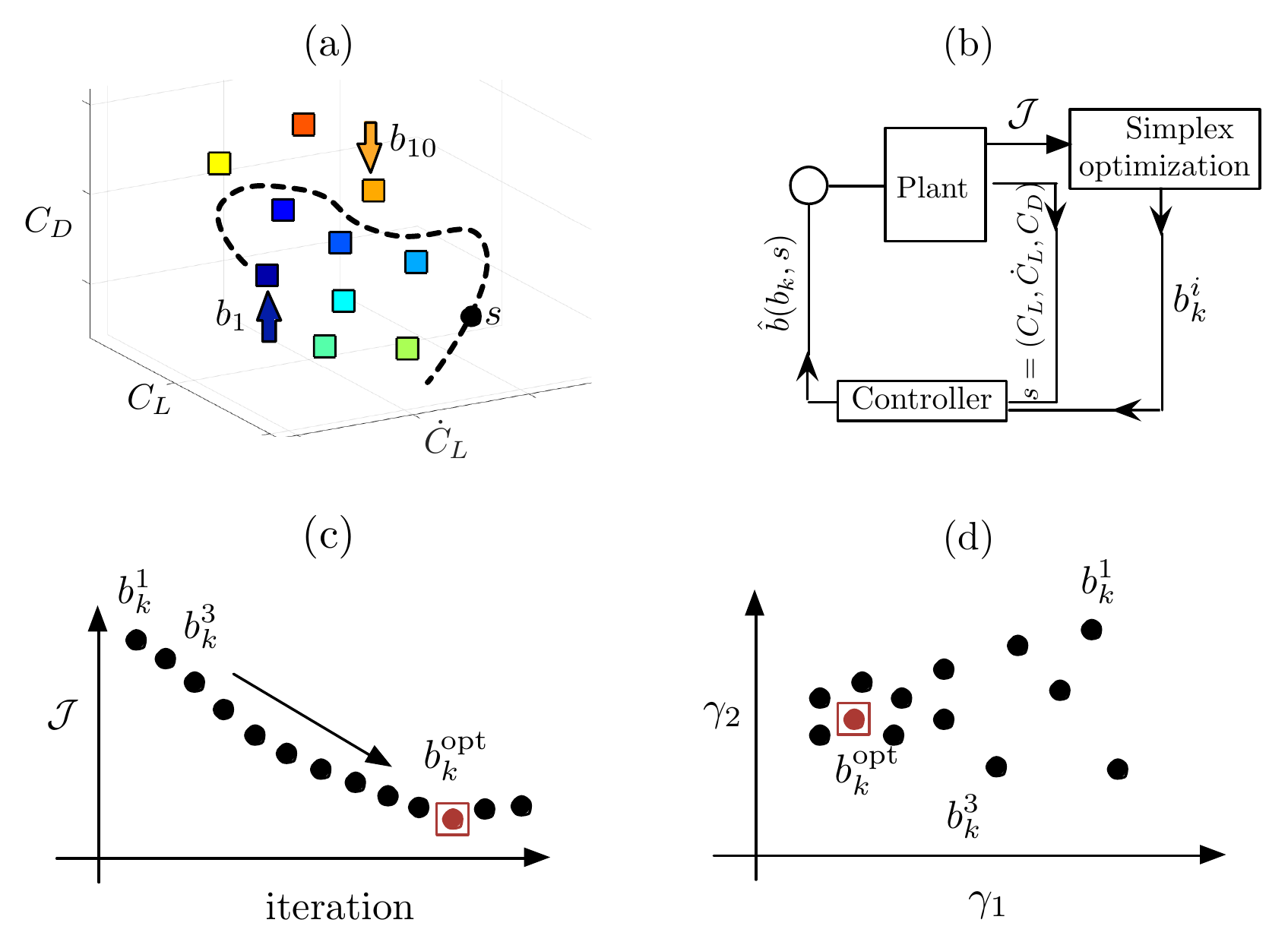} 
\end{center}
\caption{The schematic of the optimization procedure for cluster-based feedback control; (a) Control inputs assigned to cluster centroids. (b) Feedback control configuration and (c) simplex search to find optimized control law minimizing cost function $\mathcal{J}$. (d) Visualization of all the control cases on a two-dimensional proximity map $(\gamma_1,\gamma_2)$ using multi-dimensional scaling.}
\label{fig:setup3}
\end{figure} 

As opposed to these model-based approaches, we seek a feedback control law determined in a model-free manner. In particular, the segmentation of the feature space using clustering enables the parametrization of each cluster centroid with an actuation value. This is a distance-based, globally defined control law, i.e. the actuation does not neglect any dynamics. With feedback of the current state in the feature space, the central idea of the flow control strategy is to re-route the trajectories so as to maximize the performance objectives. Using an iterative optimization procedure, we can iteratively optimize the control law in an automated fashion. We now discuss the procedure for optimized cluster-based feedback control design, summarized in Figure \ref{fig:setup3}.

Each cluster $\mathcal{C}_k$ with its associated centroid $c_k$ is assigned a chosen constant control amplitude $b_k$ as shown in Figure \ref{fig:setup3}(a). This provides a blowing/suction jet velocity for each cluster. Cluster control amplitudes are then interpolated over the feature space using a normalized radial basis kernel \citep{wand1994kernel}. The current state of the observable $s(t)$ at time $t$ is used for performing feedback control. With the cluster centroids, the current sensor measurement and the cluster control amplitudes, the forcing amplitude $b$ as required in Eq. (\ref{ujet}) is provided by
\begin{equation}
b(b_k,s(t)) = \beta\frac{\sum_{k=1}^{K} b_k e^{-||s(t) - c_k ||^2/J_i}}{\sum_{k=1}^{K} e^{-||s(t) - c_k ||^2/J_i}},
\label{bhat}
\end{equation}
where $\beta$ is the feedback gain which is set to unity, unless otherwise noted. We consider higher values of $\beta$ for 3D flow control studies in \S\ref{sec:3Dresults}. The flow control is implemented as a proportional feedback controller depending on the current state in the feature space as shown in Figure \ref{fig:setup3}(b).

Iterative optimization of the cluster control amplitudes $\{b_k\}_{k=1}^K$ is then performed to minimize a cost function consisting of both state and control variables. For optimized control, we need to minimize the sum of the aerodynamic power loss ($P_\text{drag}$) and the actuation power input ($P_\text{act}$) leading to an objective function $\mathcal{J} = P_\text{drag} + P_\text{act}$. The aerodynamic power is the power required by the system to overcome drag.  Let $W$ and $V$ be the weight and speed of the flying vehicle, respectively. We can define $V = \sqrt{\frac{W}{\frac{1}{2}\rho C_L A}}$. The aerodynamic power can then be evaluated as, 
\begin{equation}
     P_\text{drag} = \frac{F_x V}{\frac{1}{2}\rho U_\infty^3 A}=  \frac{\frac{1}{2} \rho C_D V^3 A}{\frac{1}{2}\rho U_\infty^3 A} = \frac{C_D}{U_\infty^3}\left(\frac{W}{\frac{1}{2}\rho C_L A}\right)^{3/2} = \eta \frac{C_D}{C_L^{3/2}},
\label{Pa}
\end{equation}
where $\eta = \left(\frac{W}{\frac{1}{2}\rho U_\infty^2 A}\right)^{3/2}$. At cruise (steady) condition, lift is equal to the weight of the flying vehicle $W$. Maximum endurance of flight can be obtained by minimizing aerodynamic (propulsive) power. This minimum energy expenditure occurs when $C_D/C_L^{3/2}$ is minimum \citep{anderson1999aircraft}. To extract the aerodynamically favorable gain with control, we set the aerodynamic power to the baseline drag power for the unforced case by considering $\eta = \overline{C}_L^{3/2}$, where  $\overline{C}_L$ is the mean baseline lift. It must be noted that the emphasis of this work is in minimizing  the drag power and any benefit from lift force is weighed according to the scaling derived in Eq~(\ref{Pa}) to maximize flight endurance. 

The unsteady actuation power is related to the momentum injected to the fluid as
\begin{equation}
P_\text{act} =  \frac{2}{TU_\infty^3 A} \int_0^T \int_{-w/2}^{w/2}\int_{-\xi_a}^{\xi_a}|u_{jet}|^3{\rm d}\xi {\rm d}z {\rm d}t,
\label{Pb}
\end{equation}
where $T$ is the finite time horizon of application of control. The time-averaged momentum coefficient corresponding this actuation power is 
\begin{equation}
\overline{C}_\mu = \frac{2}{TU_\infty^2 A} \int_0^T \int_{-w/2}^{w/2}\int_{-\xi_a}^{\xi_a}|u_{jet}|^2{\rm d}\xi {\rm d}z {\rm d}t.
\label{Cmu}
\end{equation}

To determine the optimized cluster control amplitudes, we utilize the simplex search algorithm \citep{nelder1965simplex}, which is a gradient-free multidimensional unconstrained optimization technique. The simplex method iteratively optimizes the cluster-based feedback control laws $\{b_k^{i}\}_{k=1}^K$ which is the input to the feedback controller shown in Figure \ref{fig:setup3}(b). The superscript $i$ indicates the iteration number of control case. To start the iterative optimization, an initial simplex is formed by a Latin-hypercube sampling. As our goal is to determine $K$ optimized cluster control amplitudes, we define an initial simplex of $N_b = K+1$ vertices. Each vertex of the simplex is evaluated with a controlled LES, each with a unique set of $K$ cluster control amplitudes to define $b$. We incorporate a Latin-hypercube sampling of the parametric space of $N_b \times K$ control amplitudes for the initial simplex, which leads to a near-random sample of parameter values from a multidimensional distribution \citep{mckay2000comparison}.

We simulate the $N_b$ initial flow control cases over a finite-time horizon $T$. This time horizon is chosen as multiples of characteristic time period, derived from the shedding frequency $St = fL_c\sin(\alpha)/U_\infty$ of the flow. Once the initial $N_b$ control cases are simulated and the objective function $\mathcal{J}$ for each case is evaluated, the simplex search algorithm performs reflection, expansion, contraction and shrinkage on the cluster control amplitudes to minimize the objective function $\mathcal{J}$. These operations quickly span the search space of cluster amplitudes to find the optimized control law. The optimization procedure is terminated when the standard deviation of the currently evaluated simplex is less than a set tolerance of $\epsilon = 0.004$. Ultimately, a choice of cluster control amplitudes $\{b_k^\text{opt}\}_{k=1}^K$ results in a minimal $\mathcal{J}$, maximizing flight endurance. We call this case as the optimized control case based on the set tolerance. A schematic of simplex optimization is shown in Figure \ref{fig:setup3}(c).

As we have $K$ cluster control amplitudes to optimize, visualization of the control landscape over all the clusters can be very insightful but also challenging. Multidimensional scaling (MDS) visualizes the organization of high-dimensional objects by finding a low-dimensional subspace, which optimally preserves the distances between objects in the high-dimensional space \citep{young1938discussion}. Here, we employ MDS to visualize the similarity between control laws by finding a low-dimensional embedding maintaining pairwise distances between them~\citep{kaiser2017cluster,kaiser2017control}. In addition to measuring the similarity (or dissimilarity) between the evaluated control laws, the visualization helps in tracking of the search directions tending towards the optimum of our objective. The pairwise distances in MDS are defined as
\begin{equation}
D_{ij} = \sqrt{\frac{1}{2}\sum_{t=1}^{T} [b(b^i_k,s^i(t)) - b(b^j_k,s^i(t))]^2 + \frac{1}{2}\sum_{t=1}^T [b(b^i_k,s^j(t)) - b(b^j_k,s^j(t))]^2},
\label{Djk}
\end{equation}
where $b$ is defined by Eq.~\eqref{bhat}. The superscripts $i$ and $j$ indicate the iteration number corresponding to the control cases in the optimization procedure. MDS then aims to find a set of points $\{\gamma^i\}_{i=1}^{N_c}$, where $N_c$ is the total number of evaluated control laws, in a low-dimensional subspace such that $||\gamma^i - \gamma^j|| \approx D_{ij}$ is approximated in a least-squares sense. Here, we find the two-dimensional subspace, $\gamma^i = \{\gamma_1,\gamma_2\}$, for visualization purposes. 

To extract the two-dimensional co-ordinate subspace, we construct $B = -\frac{1}{2}CD^2C$ using a centering matrix $C = \text{I} - \frac{1}{M}1 1'$, where $M$ is the number of control cases. Here, $\text{I}$ is the identity matrix and $1$ is column array of all ones of size M. We then determine $(\gamma_1,\gamma_2) = V\Lambda^{\frac{1}{2}}$, where $\Lambda$ and $V$ contain the first two eigenvalues and eigenvectors of $B$. In general, we can find a $m-$dimensional subspace retaining the first $m$ eigenvalues and eigenvectors. This proximity map evaluates the relative performance of each control case with iterative optimization based on the variations in $\gamma_1$ and $\gamma_2$ variables, which capture the two directions in the identified subspace. Moreover, these proximity maps can accelerate the optimization process by estimating the performance of untested control laws as proposed by \cite{kaiser2017control}.

Using the above approach, the optimized cluster-based feedback control law can be determined. Unlike other control strategies which require either the linearized Navier--Stokes equations or reduced-order models to develop feedback control laws, the present framework provides a model-free alternatives. Instead of performing adjoint computations or tuning parameters associated with PID controllers, the iterative optimization strategy combined with cluster-based analysis deduces optimized control laws in a purely data-based manner. The clustering analysis and optimization just require the information of force measurements as inputs to the algorithms. Thus, the approach is easily extendable to experiments.

We demonstrate the optimization of cluster control amplitudes for the 2D separated flow in the following section. We then extend the approach to find cluster-based control laws for 3D separated flows, for which the associated computational cost and complexity of flow control is manageable but increases significantly.    

\section{Control of 2D separated flow over an airfoil}
\label{sec:2Dresults}

In this section, we demonstrate the cluster-based feedback control optimization for a 2D flow over an airfoil to maximize performance. In particular, we first present the clustering results, based on data from the baseline flow in Sec.~\ref{subsec:2Dbaseline}, that are employed to partition the feature space and provides the foundation for optimizing the control laws. We then demonstrate how the coarse-grained control law is optimized in Sec.~\ref{subsec:2Dcontrol}. The iterative optimization procedure is further analyzed using proximity maps. The resulting change in dynamics with control is examined using Markov transition models.   

\subsection{Baseline feature space clustering}
\label{subsec:2Dbaseline}

\begin{figure}
\begin{center}
\includegraphics[width = 0.99\textwidth]{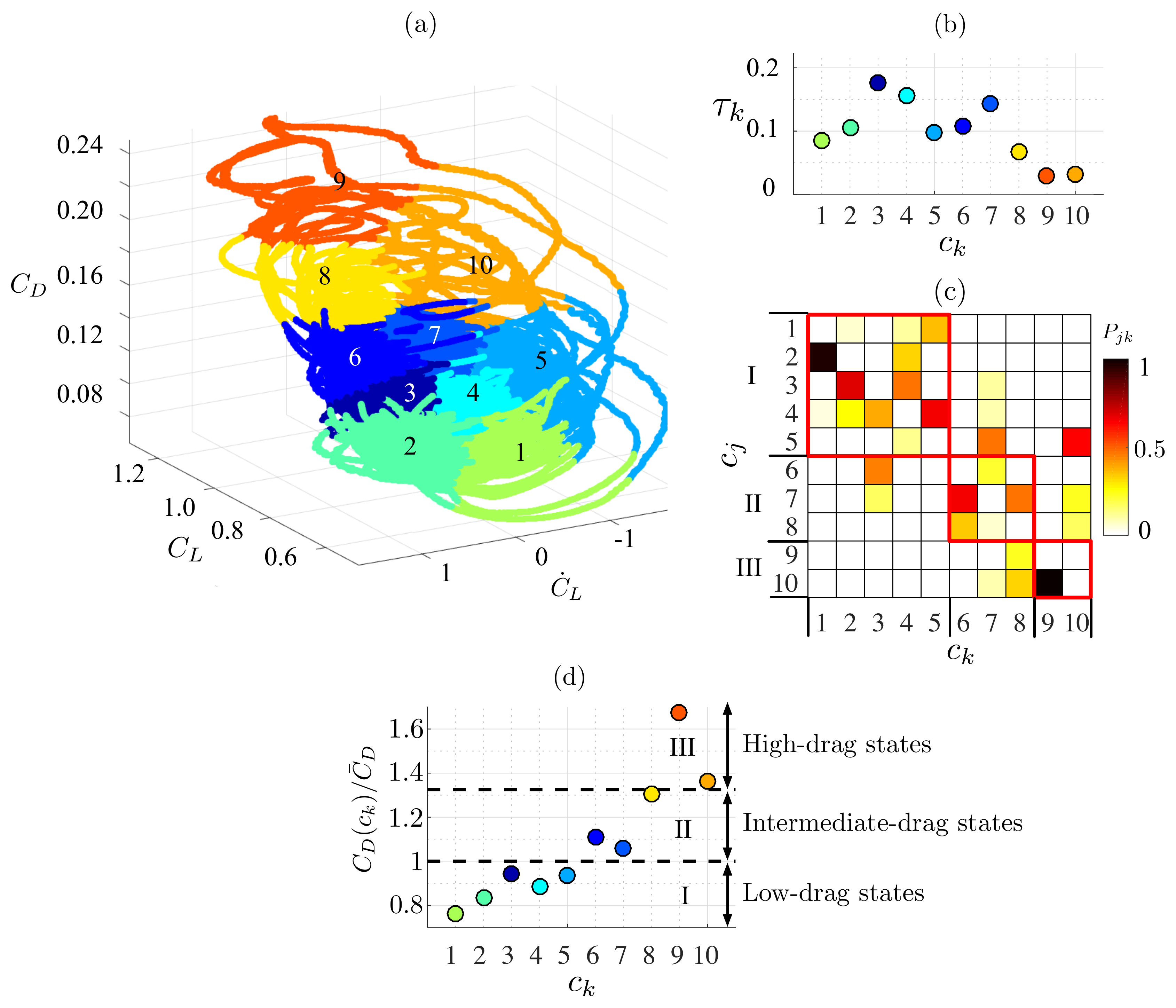} 
\end{center}
\caption{(a) Feature space clustering of the 2D baseline flow. (b) Cluster residence probabilities $\tau_k$. (c) Cluster transition probability matrix, with the red boxes indicating cluster subsets. (d) Normalized drag corresponding to cluster centroids.}
\label{fig:2Dclustering}
\end{figure}

A cluster-based analysis of the 2D baseline separated flow over the airfoil is performed as described in \S \ref{form:cbd}. In post-stalled configurations, a strong adverse pressure gradient due to separation causes a large increase in the associated pressure drag, thereby increasing the pressure losses and enlarging the size of the wake. Using a cluster-based analysis, we identify characteristic phases of the flow associated with these losses. For the analysis, time-series data of baseline trajectories $s^b(t)$ are collected. A constant time step $\triangle t$ is chosen so as to resolve the baseline vortex shedding frequency, $St = fL_c\sin(\alpha)/U_\infty = 0.081$ with at least $500$ snapshots. The collected data spans a total convective time of $tL_c/U_\infty = 110$. The data of the observable $s^b(t)$ is then partitioned into $K = 10$ clusters using the {\it k-means} algorithm as shown in Figure \ref{fig:2Dclustering}(a). The cluster discretization of the feature space separates the characteristic phase regimes of the flow providing a representative phase map. The stochasticity in the dynamics can be observed from the trajectories in the feature space. 

By analyzing the transitions between clusters, transition dynamics between characteristic phases of the flow can be deduced. These transition dynamics can be described by Markov chains and underlie the cluster-based reduced-order modeling (CROM) framework \citep{kaiser2013cluster}. The Markov model associated with CROM provides a stochastic framework for transition dynamics and its analysis can be used to identify different, weakly connected dynamic regimes. Here, the model is used to extract corresponding transition dynamics associated with post-stalled flows. The resulting CROM model describes the evolution of probability densities in state space \citep{lasota2013chaos,bollt2013applied}. The cluster-based model is represented by a probabilistic cluster transition matrix $P$. The elements of this matrix describe the probability of transition from cluster $c_k$ to $c_j$ in one forward time step $\triangle t$ and are given as 
\begin{equation}
P_{jk} = \frac{N_{jk}}{N_k},
\label{Pjk}
\end{equation}
where $N_{jk}$ is the number of transitions from cluster $c_k$ to $c_j$. The cluster transition matrix has a property of $\sum_j P_{jk} = 1$. The diagonal elements of this matrix represent the residence likelihood within the same cluster and the off-diagonal entries represent the inter-cluster transitions. We further define the probability of the flow to reside in one of the clusters, which also amounts to the average time the flow remains in any of the clusters, as $\tau_k = |N_k|/N$, also referred to as the cluster probability. The cluster probabilities and the cluster transitions are shown in Figures \ref{fig:2Dclustering}(b) and (c), respectively. The cluster probability is the highest for cluster $3$ followed by clusters $4$ and $7$. The flow states in clusters $1$ and $9$ always transition to clusters $2$ and $10$ respectively, resulting in a high probability of transition. 

To further simplify these transitions, we can identify the groups of clusters called cluster subsets, in which transitions occur more frequently within them than between them. To identify the partition of these decomposable sub-Markov chains \citep{kontovasilis1995markov}, we use the directed modularity maximization algorithm \citep{leicht2008community}. The algorithm detects modular subsets where the transitions between clusters within a subset are dense compared to the transition between clusters from different subsets. Here, the algorithm identifies three subsets: subset I:$\{1,2,3,4,5\}$, subset II:$\{6,7,8\}$, and subset III:$\{9,10\}$. These subsets are identified by the red boxes in Figure \ref{fig:2Dclustering}(b). High probability inter-subset transitions are observed between clusters $3\rightarrow 6$, $7\rightarrow 5$ and $10 \rightarrow 5$. These transitions also mark the major paths of transition between subsets. Most of the remaining transitions are within subsets. We further analyze the drag contribution of each cluster, which is obtained from the cluster centroids. We normalize the drag coefficient associated with each cluster centroid with the mean drag $\overline{C}_D$ as shown in Figure \ref{fig:2Dclustering}(d). The dashed lines separate the subsets. Interestingly, subset I corresponds to low-drag states of the flow $C_D/\overline{C}_D \lesssim 1$, subset II corresponds to intermediate drag states $1\lesssim C_D/\overline{C}_D \lesssim 1.325$ and subset III corresponds to the high-drag states $C_D/\overline{C}_D \gtrsim 1.325$. Thus, different levels of drag are all dynamically separated, which can significantly simplify the control design. These insights suggest steering the flow to the subset associated with the lowest drag and then keeping the state in the low-drag subset.

\begin{figure}
\begin{center}
\includegraphics[width = 0.99\textwidth]{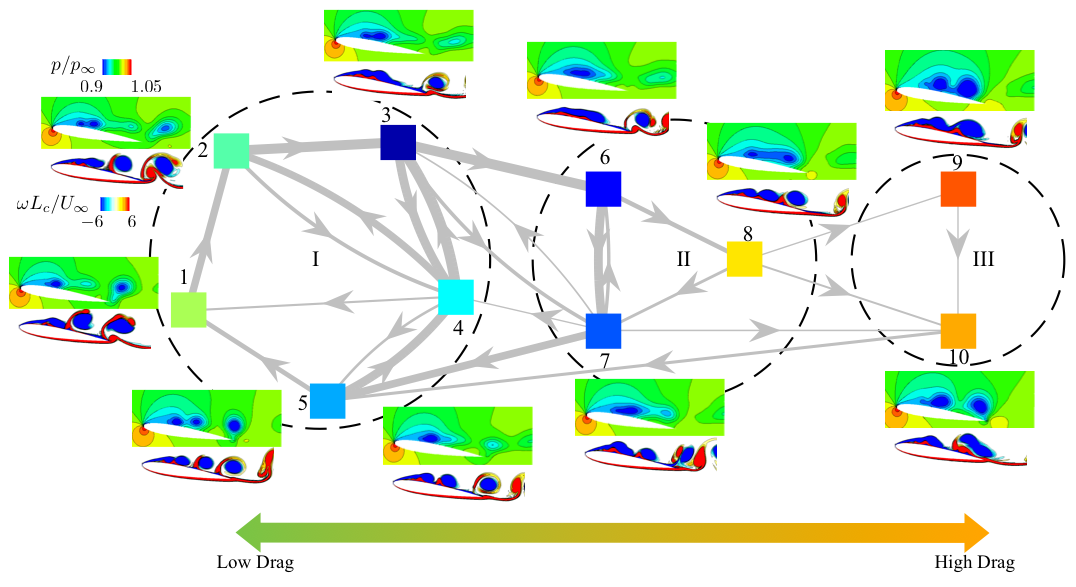} 
\end{center}
\caption{Graph of Markov chain highlighting transitions between clusters for 2D separated flow over a NACA0012 airfoil. 2D cluster-averaged pressure flow fields ($p/p_\infty$) and instantaneous vorticity contours ($\omega L_c/U_\infty$) corresponding to each cluster are shown. The dashed circles indicate the cluster subsets.}
\label{fig:2DMclustering}
\end{figure}

An alternate view point of the Markov chain is a random walk on a directed graph with nodes being clusters and edges being the transition dynamics between them \citep{Newman10}. The edge weights correspond to the number of transitions between the respective clusters $N_{jk}$ normalized by the maximum number of cluster transitions observed in the entire trajectory, $\max(N_{jk})$. We neglect self-loops, i.e. transitions within the cluster and edges weights less that a threshold of $0.1$. Self-loops are related to the cluster probabilities and edge weights less than the given threshold do not contribute significantly to the overall transition dynamics. The directed graph representation of the cluster transitions is shown in Figure \ref{fig:2DMclustering}. The edge weights are indicated by the thickness of the lines for this network visualization. The subsets are also shown in the figure. Here, we can clearly identify the paths of cluster transitions and visualize the phase evolution of the flow. One characteristic feature that stands out from this graph visualization is the role of cluster $8$. The flow can reach cluster $8$ only from cluster $6$. This means that only through this cluster $8$, the flow can transition to the high-drag state of cluster $9$ or $10$. For this reason, we call cluster $8$ the switching cluster. The significance of the cluster transition pathway $6\rightarrow8\rightarrow9$ is an important consideration for flow control. The elimination or avoidance of this cluster transition pathway is key in enhancing aerodynamic performance, particularly for drag reduction. 

Understanding the flow behavior in each cluster is critical in understanding changes in flow physics leading to transitions from low-drag to high-drag states and vice versa. We can average the flow snapshots within each cluster to obtain cluster representative flow fields. The cluster-averaged pressure and the instantaneous vorticity fields are shown in Figure \ref{fig:2DMclustering}. The cluster flow fields in subset I provide an enhanced understanding of the flow physics at the low-drag states of the flow. The different phases of the flow are clearly visible in the cluster transitions $3\rightarrow4\rightarrow5\rightarrow1\rightarrow2$ upon observing the associated vortex dynamics. We see the initiation of shear-layer roll up in cluster $3$ and the presence of leading-edge vortices in cluster $4$. These vortices start shedding mid-chord in cluster $5$. In cluster $1$, we observe the trailing-edge vortex sheet roll-up followed by the von K\'{a}rm\'{a}n vortex shedding in the wake in cluster $2$. In these low-drag states, the low-pressure core above the suction side of the airfoil is closer to the leading edge, especially in clusters $1$ and $2$.

The cluster transition from $3$ to $6$ is characterized by the strengthening of the low-pressure core near the mid-chord of the airfoil. This results from the elongation of the vortex sheet on the airfoil surface. Following the graph representation, the flow at cluster $6$ could transition either to cluster $7$ or $8$. Cluster $7$ is characterized by the shear layer roll-up and intermittent shedding in the wake. The rolled-up vortices shed near the mid-chord of the airfoil resulting in a flow transition $7\rightarrow 5$. The transition to cluster $8$ results when the rolled-up vortices grow in size elongating the pressure core. As opposed to cluster $5$, here the vortices do not detach from the airfoil surface but are arranged compactly over the surface. The flow from cluster $8$ transitions to cluster $9$, accompanied by a low-pressure core spanning the entire airfoil with a large roll-up near the trailing edge of the airfoil. This results in a fully separated flow and leads to high-drag. Following this state, the trailing-edge vortex sheet rolls up and the vortices detach from the airfoil at the trailing edge leading to a subsequent transition to cluster $10$ and then to $5$. In cluster $10$, the flow briefly reattaches and then separates near the trailing edge. Moreover, cluster $7$ is characterized by high-frequency shedding in the wake as opposed to the low-frequency shedding in cluster $10$.  

  \begin{figure}
\begin{center}
\includegraphics[width = 0.75\textwidth]{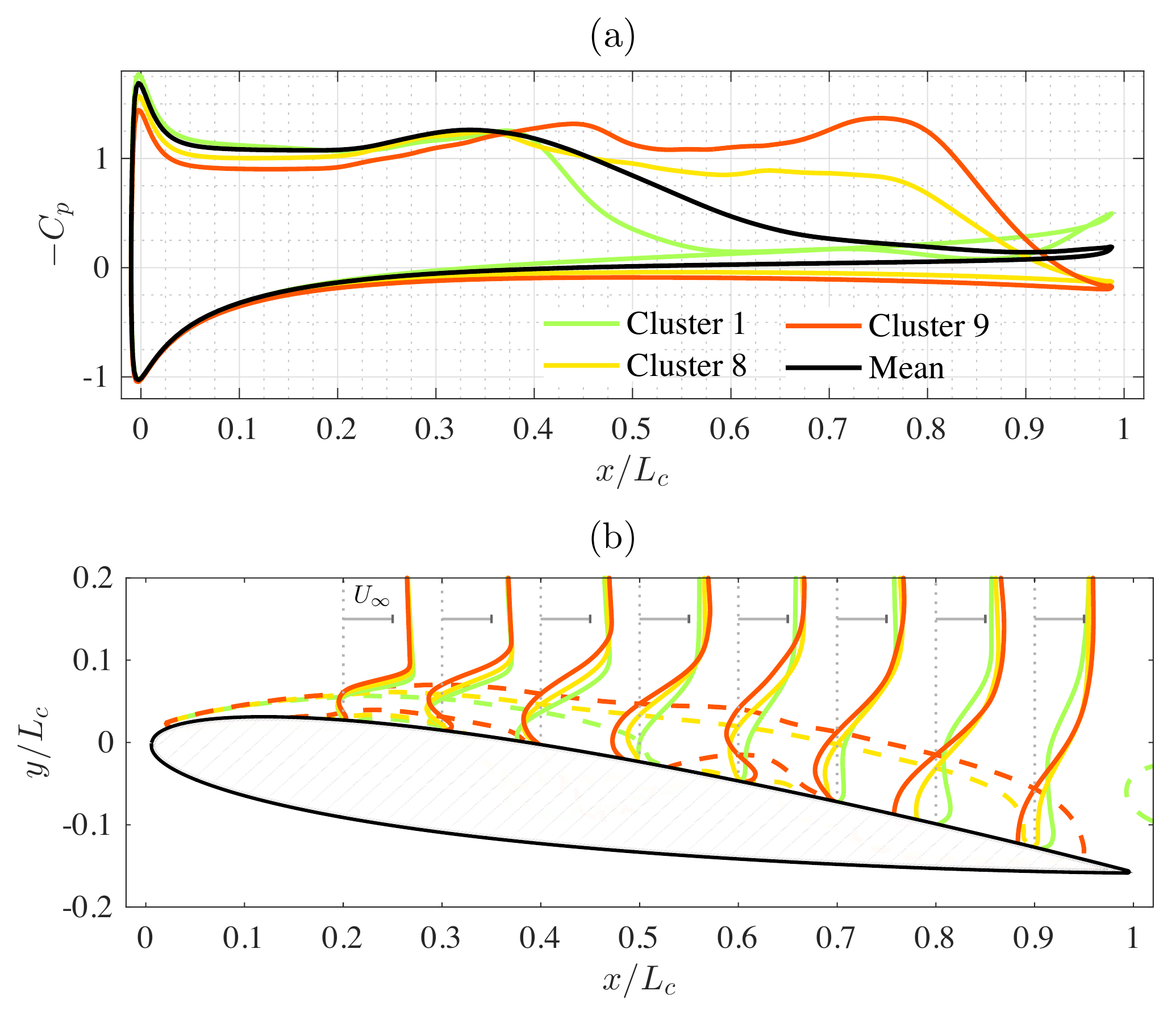} 
\end{center}
\caption{(a) Cluster-averaged pressure distribution over the suction and pressure surfaces of the airfoil and (b) cluster-averaged streamwise velocity profiles for the 2D baseline flow. Dashed lines indicate the contours of $\bar{u}_x/U_\infty = 0$. }
\label{fig:2Dclusterphysics}
\end{figure} 

The significance of the clusters can also be examined by looking at the cluster-averaged pressure distribution over the airfoil. We show the pressure distribution over the airfoil surface and streamwise velocity profile for one representative cluster in each cluster subset in Figure \ref{fig:2Dclusterphysics}(a) and (b), respectively. On the suction side of the airfoil, we can see a favorable pressure gradient near the mid-chord for cluster $1$ (low-drag cluster) with a shorter flat pressure region indicating the shorter region of separation. This is also shown by the dashed lines corresponding to zero time-averaged streamwise velocity contour, $\bar{u}_x/U_\infty = 0$. Clusters $8$ and $9$ have larger regions of separation and are associated with an adverse pressure gradient near the trailing edge.

The above discussion of the cluster transitions and associated evolution of the flow are based on discretizing the feature space into $K = 10$ clusters. The clustering analysis is based on the {\it k-means} algorithm which is primarily concerned with minimizing the total within-cluster variance. The current cluster analysis focuses on obtaining the best compressed coverage of trajectories in the feature space. Such a data-driven, state-space discretization enables us to obtain a coarse-grained global control law over the entire feature space. It must be noted that increasing the number of clusters may reveal additional flow paths and other clustering techniques may result in alternate state space discretizations.  Although there may be variability in the phases of the flow characterized with alternate discretizations, the fundamental features of the clusters transitions remain.

\subsection{Optimized Feedback control}
\label{subsec:2Dcontrol}

Based on the coarse-graining of the feature space into clusters discussed in the previous section we design a cluster-based feedback control law. The actuation input $b(t)$ in Eq. ~\eqref{bhat} at time $t$ is determined with the feedback of the observable $s(t)$ in  the feature space. The actuation input takes into account the control amplitudes at each cluster-discretized phase of the flow $\{b_k\}_{k=1}^K$ and the relative distance of the current state to the cluster centroids $\{c_k\}_{k=1}^K$ to obtain a smooth control law over all the clusters. Thus, actuation inputs associated with cluster in close proximity to the current measurement are weighed more strongly than those associated with clusters farther away. This defines a global control law based on an initial clustering of the baseline configuration and does not rely on the identification of characteristic coherent structures in the flow. 

\begin{figure}
\begin{center}
\includegraphics[width = 0.815\textwidth]{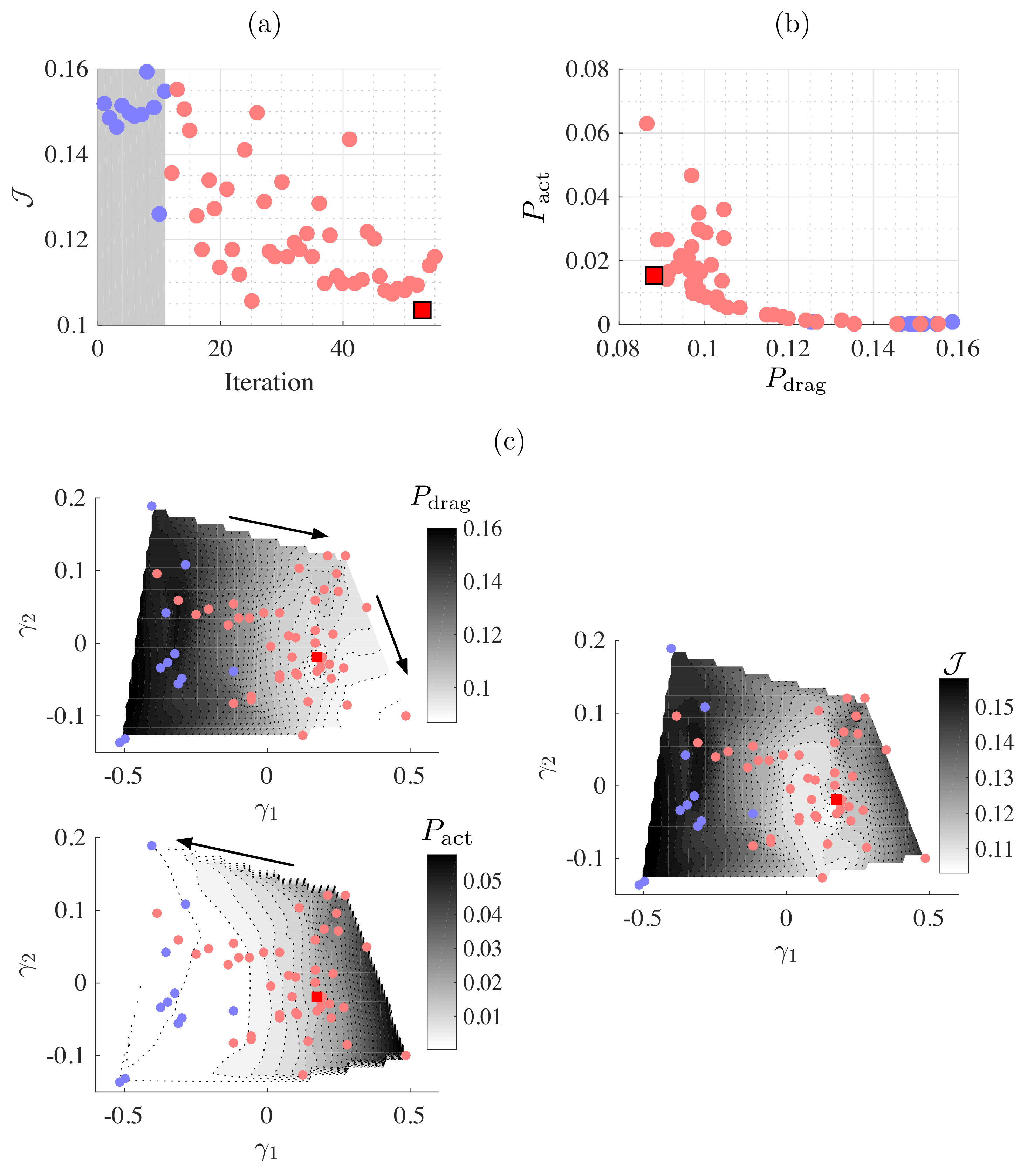} 
\end{center}
\caption{Control of 2D separated flow: (a) Objective function minimization to determine the optimized control law and (b) Power consumption. The red square symbol denotes the optimized case.  (c) Control objective landscape, $P_\text{drag} = P_\text{drag}(\gamma_1,\gamma_2)$, $P_\text{act} = P_\text{act}(\gamma_1,\gamma_2)$, and $\mathcal{J} = \mathcal{J}(\gamma_1,\gamma_2)$ balancing $P_\text{drag}$ and $P_\text{act}$, all three determined using multidimensional scaling. The transparent blue dots indicate initial simplex control cases. The arrows indicate directions of minimization.}
\label{fig:2Dobjective}
\end{figure}

The control amplitudes $\{b_k\}_{k=1}^K$ are then iteratively optimized using a simplex search to achieve the desired objective as discussed in \S \ref{form:ofcd}. We design an initial simplex using Latin hypercube sampling with zero-mean offset control amplitudes $\sum_k b_k = 0$. As the forcing amplitude, given by Eq.~\eqref{bhat}, is a weighted summation over cluster control amplitudes and depends on the distance from the cluster centroids, this zero-mean offset is not be guaranteed for feedback control. Time averages in controlled flow are estimated over 12 periods of the baseline shedding frequency. The optimization of the cost function $\mathcal{J}$ is summarized in Figure \ref{fig:2Dobjective}(a). Following the initial simplex consisting of $K+1$ control cases (indicated by blue dots), cost functional is minimized iteratively. The square symbol denotes the optimized control case. The optimization over aerodynamic power $P_\text{drag}$ and actuation power $P_\text{act}$ is shown in Figure \ref{fig:2Dobjective}(b). 

To visualize the control landscape, MDS is performed, as discussed in \S\ref{form:ofcd}. MDS identifies a low-dimensional embedding of the control cases in terms of a proximity map. Here, we extract a 2D proximity map over the ($\gamma_1, \gamma_2$) space. Each point in this proximity map stands for a control case or control law, respectively. Pairwise distances given by Eq.~\eqref{Djk} measure the similarity/dissimilarity between the control cases. The similarity between control laws in this map increases as the distance between them gets smaller. We fit surfaces of the form $P_\text{drag} = P_\text{drag}(\gamma_1,\gamma_2)$, $P_\text{act} = P_\text{act}(\gamma_1,\gamma_2)$ and $\mathcal{J} = \mathcal{J}(\gamma_1,\gamma_2)$ to all the evaluated control laws in Figure \ref{fig:2Dobjective}(c). The proximity maps provide information on the complexity of the objective functions, e.g. a single vs. multiple minima, and indicate optimization directions for minimizing the aerodynamic and actuation power. Minimizing aerodynamic power is associated with the direction of increasing $\gamma_1$, while minimizing actuation power is associated with direction of decreasing $\gamma_1$. As actuation power input increases (increasing $\gamma_1$), the aerodynamic performance improves. For $\gamma_1>0$, minimal aerodynamic power is obtained with decreasing $\gamma_2$. In summary, minimizing aerodynamic power involves maximizing $\gamma_1$ and minimizing $\gamma_2$, while minimizing actuation power involves minimizing $\gamma_1$.  Balancing both power considerations, the control landscape shrinks at the optimal location for the cost function $\mathcal{J}$. Beyond the analysis of the optimization procedure, these proximity maps can help accelerate the control law optimization by estimating the expected performance of control laws without evaluating them, which can then, e.g., be discarded if a control law with a similar performance has already been evaluated. 

 \begin{figure}
\begin{center}
\includegraphics[width = 0.85\textwidth]{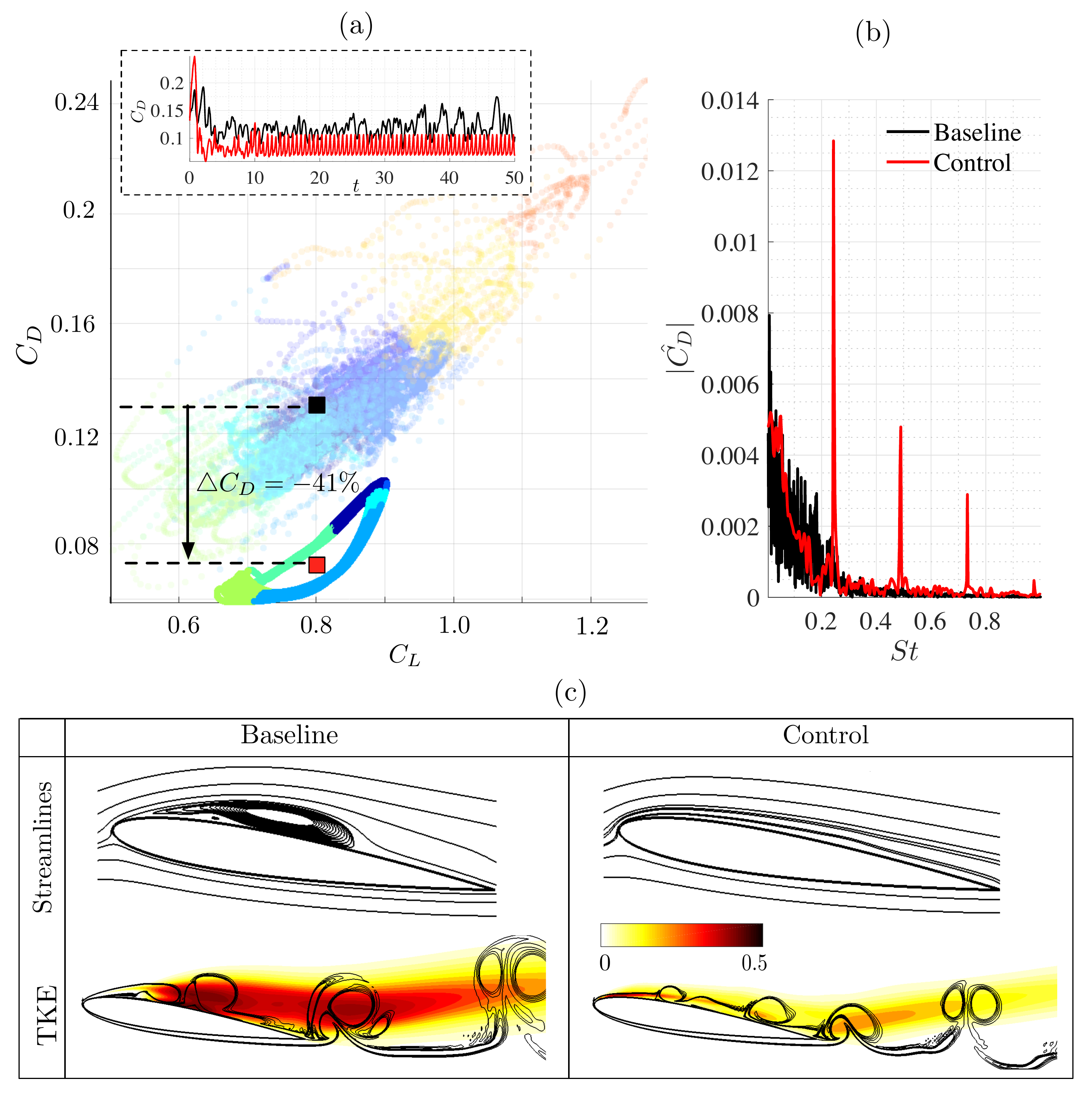} 
\end{center}
\caption{Comparison of baseline and optimized 2D flow control case: (a) Trajectories, (b) spectral analysis of drag data, and (c) time-averaged streamlines and turbulent kinetic energy (TKE) fluctuations. Baseline trajectories are shown in transparent in (a). The trajectories are colored according to their member clusters. The contour lines in (c) (bottom) indicate the instantaneous vorticity fields and the colorbar indicate the TKE fluctuation levels.}
\label{fig:2Dtrajectories}
\end{figure}

We now investigate the optimized control case to gain insights on the control strategy uncovered by cluster-based control optimization. The baseline and optimally controlled trajectories in the lift-drag coefficient plane are shown in Figure \ref{fig:2Dtrajectories}(a). A $41\%$ drag reduction is achieved with the  optimized control law. In addition, the unsteadiness in the flow is reduced. In the inset of Figure \ref{fig:2Dtrajectories}(a), the time evolution of the drag coefficient with and without control is shown. In the optimally controlled flow, the trajectories are pushed away from the high-drag states towards the low-drag states of the flow. We compare the drag spectra with single sided amplitude $|\hat{C}_D|$ in baseline and control cases in Figure \ref{fig:2Dtrajectories}(b). The dominant peak for the optimal control is obtained corresponding to a forcing frequency of $St^{+} = 0.243$, identified by the cluster-based optimization procedure. This frequency is thrice the dominant baseline shedding frequency. As the feature space is discretized based on the baseline trajectories, the cluster-based procedure identifies optimal flow actuation at harmonics of the baseline shedding frequency. The time-averaged streamlines and TKE for both the baseline and controlled flows are shown in Figure \ref{fig:2Dtrajectories}(c). The contour lines in TKE correspond to the instantaneous spanwise vorticity. The streamlines indicate that the separation bubble is eliminated with control compared to the baseline resulting in a fully attached flow. Moreover, with control, the turbulent kinetic energy fluctuation dramatically decreases. Furthermore, the roll-up of the vortices is delayed.

\begin{figure}
\begin{center}
\includegraphics[width = 0.95\textwidth]{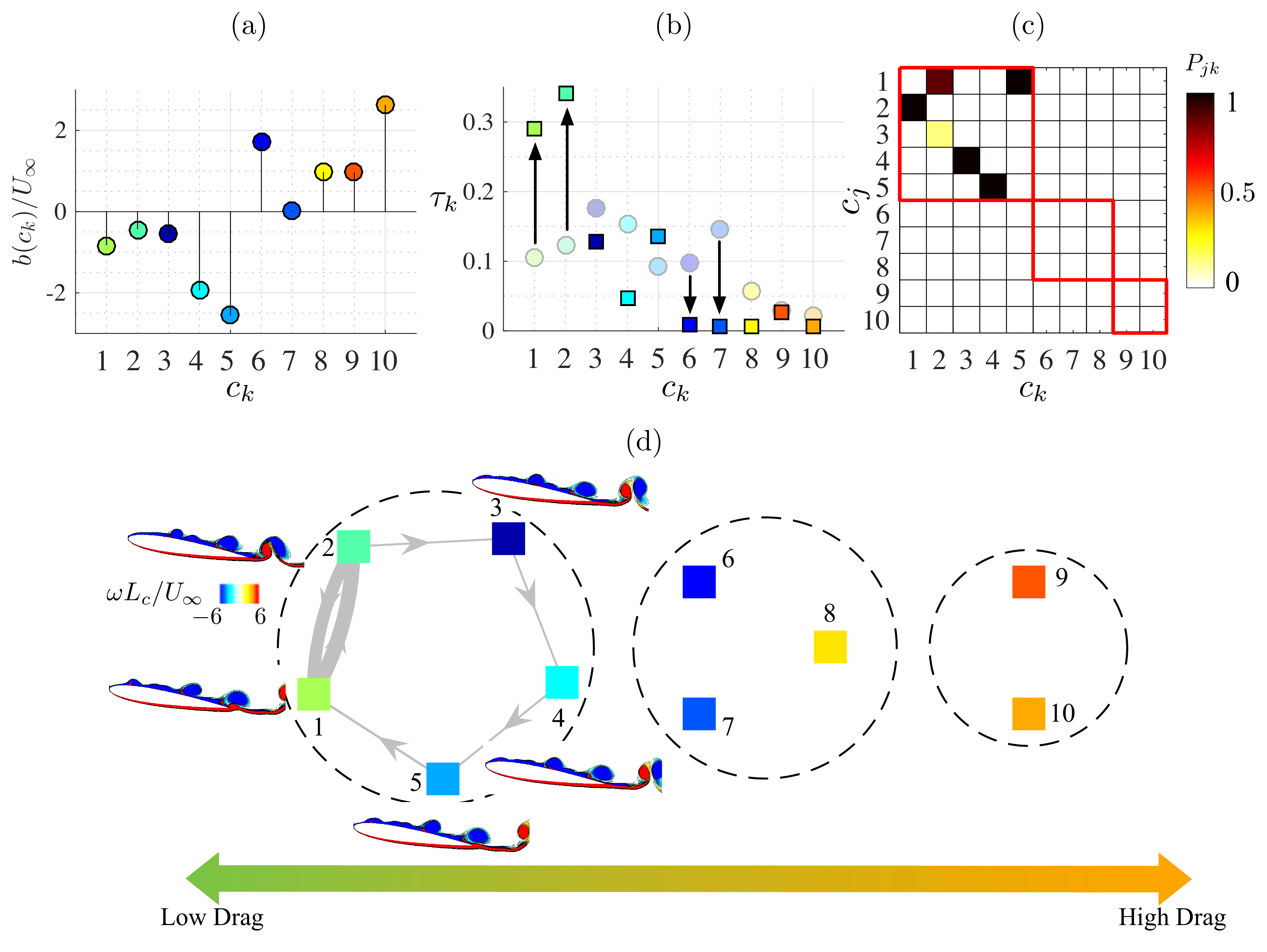} 
\end{center}
\caption{Control of 2D separated flow: (a) Cluster jet velocity, (b) cluster residence probability for baseline (shown in transparent round symbols) and controlled flows (shown in solid square symbols), (c) Cluster transitions with control, (d) controlled Markov transition network with instantaneous vorticity contours  ($\omega L_c/U_\infty$).}
\label{fig:2Dtempnetwork}
\end{figure} 

The cluster control amplitudes $b_k$ associated with the optimized control case are shown in Figure \ref{fig:2Dtempnetwork}(a). Negative and positive amplitudes indicate suction and blowing, respectively. The cluster control amplitudes are negative for clusters $1-5$ and positive for $6-10$. This indicates that suction is performed in subset I (clusters $1-5$) where $C_D/\overline{C}_D<1$. In the remaining clusters (subset II and III), blowing is performed. In the optimized control case, suction is performed in subset I to keep the trajectories in the low-drag state. As discussed in \S\ref{subsec:2Dbaseline}, the transition to high-drag cluster $9$ occurs via the path $6\rightarrow8\rightarrow9$. Blowing is performed for these clusters to kick the trajectories away from these high-drag states. For these clusters, $C_D/\overline{C}_D>1$. The flow states in cluster $9$ always transition to cluster $10$. The high blowing ratio in cluster $10$ causes the flow to transition to cluster $5$, where highest level of suction is applied to keep the flow in the low-drag states. At steady state, the mean cluster-based control amplitude is $|\bar{b}|/U_\infty = 0.78$ which amounts to a momentum coefficient $\overline{C}_\mu = 0.016$.

We evaluate the cluster probability for the optimally controlled flow and compare it with the baseline flow as shown in Figure \ref{fig:2Dtempnetwork}(b). The controlled flow spends a significant amount of time in the lowest states of clusters $1$ and $2$. The cluster probability associated with subset II clusters $6,7,8$ reduces considerably. The cluster transitions in the controlled flow and the controlled Markov transition network are shown in Figure \ref{fig:2Dtempnetwork}(c) and (d), respectively. The iterative optimization procedure coupled with cluster-based control laws achieve a re-routing of the trajectories to reduce drag power associated with the flow. This minimizes the transition to clusters in subset II (intermediate-drag states) which prevents cluster transitions to clusters in subset III (high-drag states). The controlled flow exhibits a limit-cycle behavior, which is resolved by the low-drag clusters in subset I. As the flow transitions only in subset I clusters, the dominant frequency in controlled flow increases as mentioned before. 

In summary, the cluster-based control strategy iteratively identifies optimal forcing amplitudes at the cluster states that result in minimizing power consumption for flight. For the optimized feedback control law for 2D separated flow, suction is performed for the low-drag clusters $C_D/\overline{C}_D<1$, while blowing is performed for the remaining clusters. With control, the Markov transition network is modified so that the transition dynamics shift towards the low-drag states. Thus, the flow transition to the switching cluster is avoided resulting in higher cluster probabilities associated with the low-drag clusters. From the physics standpoint, the optimized control case yields in fully attached flow leading to a drag reduction of $41\%$ compared to the the baseline flow. Thus, the cluster-based control procedure identifies an optimized feedback control law to reduce flow separation. 

\section{Control of 3D separated flow over an airfoil}
\label{sec:3Dresults}

In the previous section, the cluster-based methodology was demonstrated for 2D flow control. In this section, we extend the approach to deduce global control laws for 3D separated flows over an airfoil in a model-free manner. Despite the 3D LES computations being very expensive and flow physics being rich and complex, the present approach utilizes a low-dimensional feature space to cluster the dynamics enabling a computationally tractable flow control strategy. Using this cluster-based strategy, we deduce an optimized global feedback control law for unsteady blowing to minimize power consumption of aerodynamic flight.      

\subsection{Baseline feature space clustering}
\label{subsec:3Dbaseline}

\begin{figure}
\begin{center}
\includegraphics[width = 0.99\textwidth]{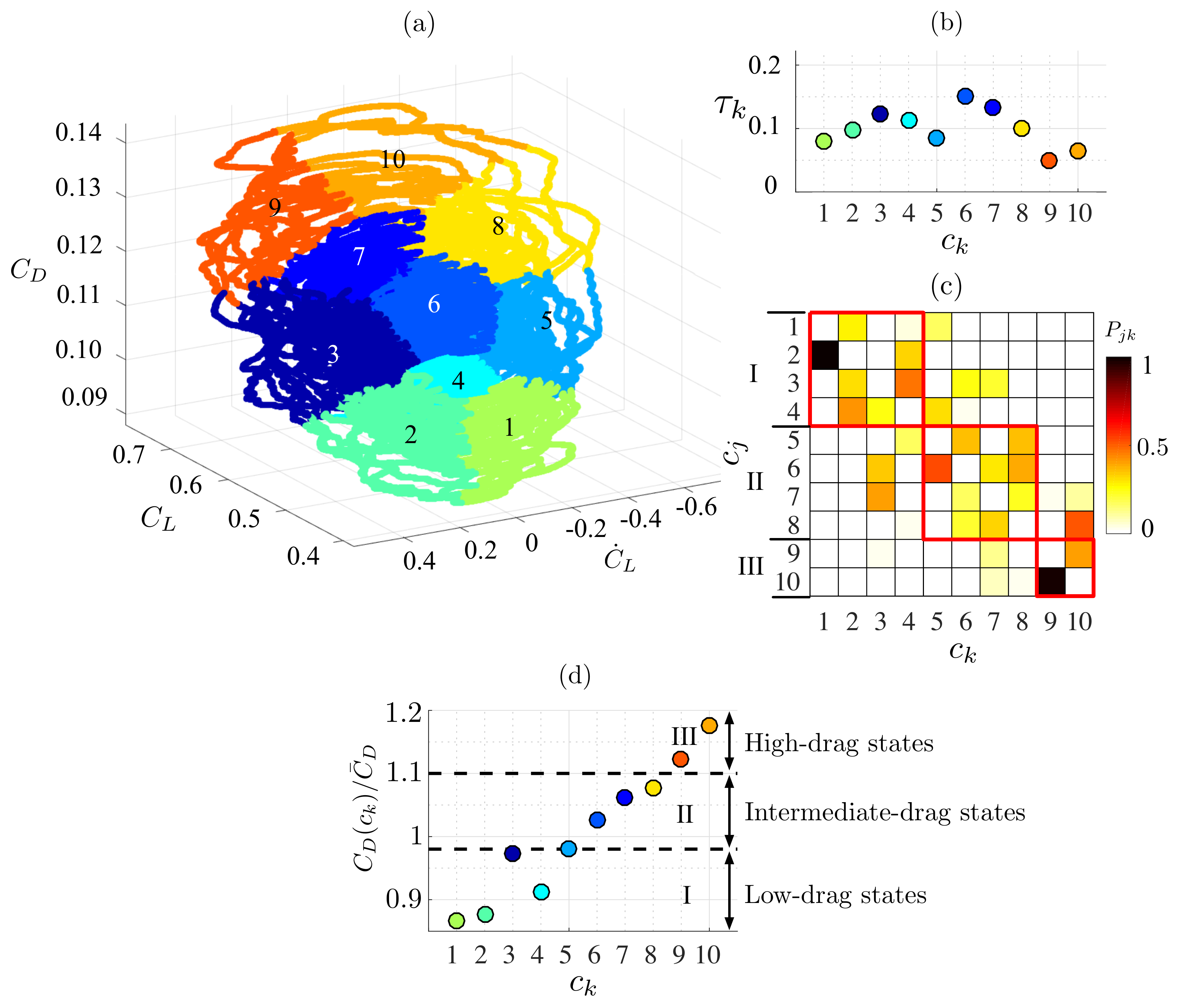} 
\end{center}
\caption{(a) Feature space clustering of 3D baseline flow data. (b) Cluster probabilities. (c) Transitions probabilities, (d) Normalized, average drag coefficient across clusters. The red boxes and the dashed circles indicate the cluster subsets.}
\label{fig:3Dclustering}
\end{figure} 

The 3D baseline separated flow over an airfoil contains flow features with von K\'{a}rm\'{a}n vortex shedding in the wake and Kelvin-Helmholtz instabilities in the shear layer. The dominant frequency associated with von K\'{a}rm\'{a}n vortex shedding is $St = 0.0884$ and the dominant shear-layer frequency associated with the Kelvin-Helmholtz instability is almost an order of magnitude higher at $St = 0.6952$. For the cluster-based analysis of the 3D baseline flow, a time series trajectory of the observables $s^b(t)$ is collected at a constant time step of $\triangle t = 0.0036$. 

The feature space segmentation into $K = 10$ clusters is  shown in Figure \ref{fig:3Dclustering}(a). For the 3D baseline flow, the range of fluctuations in $\dot{C}_L$ is less compared to the 2D baseline flow and shows more frequent cluster transitions. The variance in the cluster probabilities is reduced compared to the 2D clusters, as shown in Figure \ref{fig:3Dclustering}(b). As cluster $6$ is positioned closest to the centroid of the full data set, $\bar{c}$, its cluster probability is the highest. The transition probabilities of the cluster transition matrix are shown in Figure \ref{fig:3Dclustering}(c). The flow states in clusters $1$ and $9$ always transition to clusters $2$ and $10$, respectively, resulting in a high probability of transition. 

Performing directed modularity maximization, three subsets can be identified: subset I:$\{1,2,3,4\}$, subset II:$\{5,6,7,8\}$, and subset III:$\{9,10\}$. These subsets are highlighted in the red boxes in Figure \ref{fig:3Dclustering}(c). High probability inter-subset transitions originate from cluster $3$ in subset I, $7$ in subset II and $10$ in subset III. The grouping of the clusters into subsets can also be correlated with the drag co-ordinate, shown in Figure \ref{fig:3Dclustering}(d). Here, subset I corresponds to low-drag states of the flow $C_D (c_k)/\overline{C}_D \lesssim 0.98$, subset II correspond to intermediate-drag states $0.98 \lesssim C_D(c_k)/\overline{C}_D \lesssim 1.1$, and subset III corresponds to the high-drag states $C_D(c_k)/\overline{C}_D \gtrsim1.1$.

\begin{figure}
\begin{center}
\includegraphics[width = 0.99\textwidth]{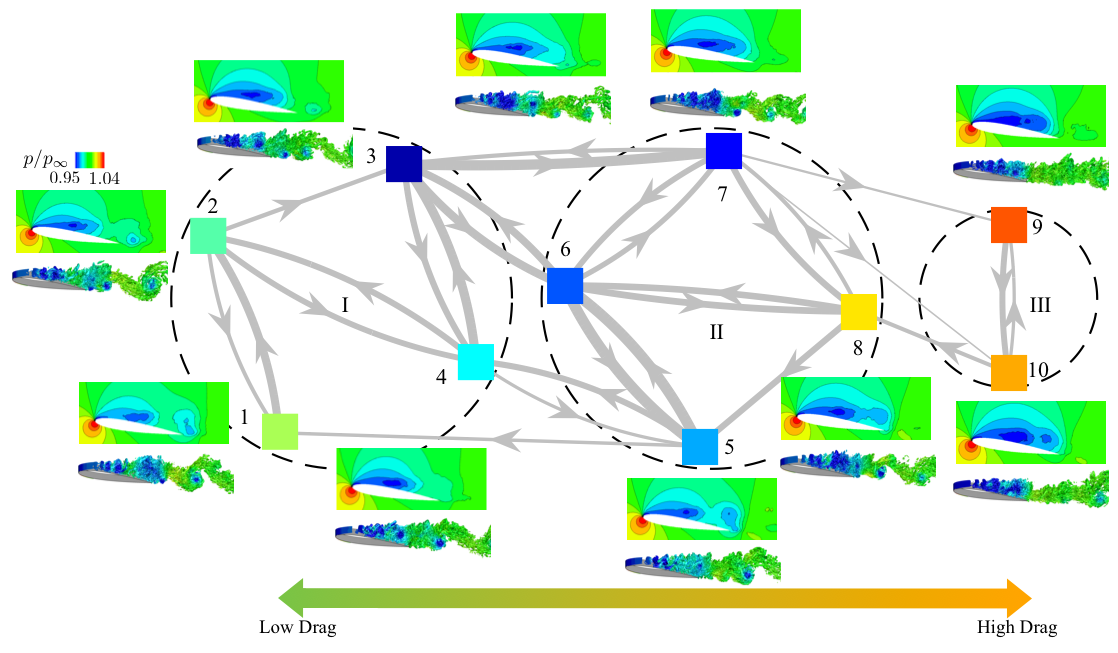} 
\end{center}
\caption{Graph of Markov chain illustrating transitions between clusters for the 3D baseline flow. 3D cluster-averaged mid-span pressure flow fields ($p/p_\infty$) and instantaneous pressure contours (highlighted by Q-criterion) corresponding to each cluster are shown. The dashed circles indicate the cluster subsets.}
\label{fig:3Dclustering2}
\end{figure} 

Further clarity in the transitions can be obtained by examining at the Markov transition network shown in Figure \ref{fig:3Dclustering2}. We notice a high volume of transitions in this 3D flow network compared to the 2D Markov transition network. High-probability cluster transitions involve cluster $6$ due to its central position in feature space. The transition to high-drag clusters $9$ and $10$ are only possible via cluster $7$, which we refer to as the switching cluster in the 3D flow configuration. The midspan pressure flow fields associated with the cluster centroids are shown in Figure \ref{fig:3Dclustering2}. We have also shown representative instantaneous flow fields with {\it Q}-criterion isosurfaces, which are colored by the pressure distribution in each cluster. Coherent structures in the wake from von K\'{a}rm\'{a}n vortex shedding in the low and intermediate drag clusters can be observed from the corresponding flow fields. The elongation of the low-pressure core near the mid-chord of the airfoil along with the associated leading-edge vortex sheet in cluster $3$ results in a transition to the switching cluster $7$, from which the flow evolves to the high-drag states. These high-drag states are accompanied by a spanning of the low-pressure core across the entire airfoil surface leading to a complete flow separation. Here, large roll-up of the vortical structures are observed near the trailing edge which do not detach. Alternatively, if the vortex sheet rolls up and detaches in cluster $3$, the resulting transitions $3\rightarrow6\rightarrow5$ ultimately lead the trajectories to low-drag clusters $1$ and $4$. Thus, the modification of the cluster pathway $3\rightarrow7\rightarrow9\rightarrow10$  is an important consideration for flow control, particularly for drag reduction. In the next section, we use the cluster partition to design the flow control strategy.

\subsection{Optimized Feedback control}
\label{subsec:3Dcontrol}

We perform feedback control using the discretized clusters in 3D baseline flow. The objective is to deduce blowing amplitudes in characteristic phases of the flow in order to maximize the aerodynamic performance. This is achieved by iteratively optimizing the control amplitudes $b_k$ in each cluster in an automated fashion to minimize the cost function $\mathcal{J}$, comprised of the aerodynamic and actuation power. In the review by \cite{Greenblatt:PAS00}, it was shown that the excitation of Kelvin-Helmholtz instabilities in the shear layer is essential for suppression of flow separation. Knowledge of these instabilities can be leveraged to design flow control strategies. Here, a cluster-based control law is optimized without assuming any prior knowledge of instabilities. The value associated  with $\eta$ in Eq.~\eqref{Pa} influences the relative importance of actuation power (input cost) and aerodynamic power (state cost) in the objective function evaluation. Increasing $\eta$ lowers the relative importance of actuation power and may yield an optimized control law  associated with higher $\overline{C}_\mu$. The feedback gain $\beta$ in Eq.~\eqref{bhat} associated with these optimized control amplitudes is subsequently increased to explore the flow control implications at higher $\overline{C}_\mu$.

To speed up the 3D computations, we employ a parallel simplex method following  \cite{lee2007parallel}. The method is similar to the original simplex method, except that multiple control simulations can be performed in parallel to accelerate the optimization process. In the 2D control effort, the cluster-based control optimization was unconstrained allowing for both blowing and suction jet velocities in the clusters. However, in the 3D control effort, a constraint is added to restrict the control amplitudes in the simplex search \citep{luersen2004constrained} such that pure blowing is performed with $0\le b_k/U_\infty \le3.3$. The lower constraint ensures that only pure blowing is performed and the upper constraint limits the highest blowing ratio that can be achieved. The addition of this constraint is motivated primarily to examine if the cluster-based strategy can take advantage of flow instabilities for the control of separated flows.

\begin{figure}
\begin{center}
\includegraphics[width = 0.825\textwidth]{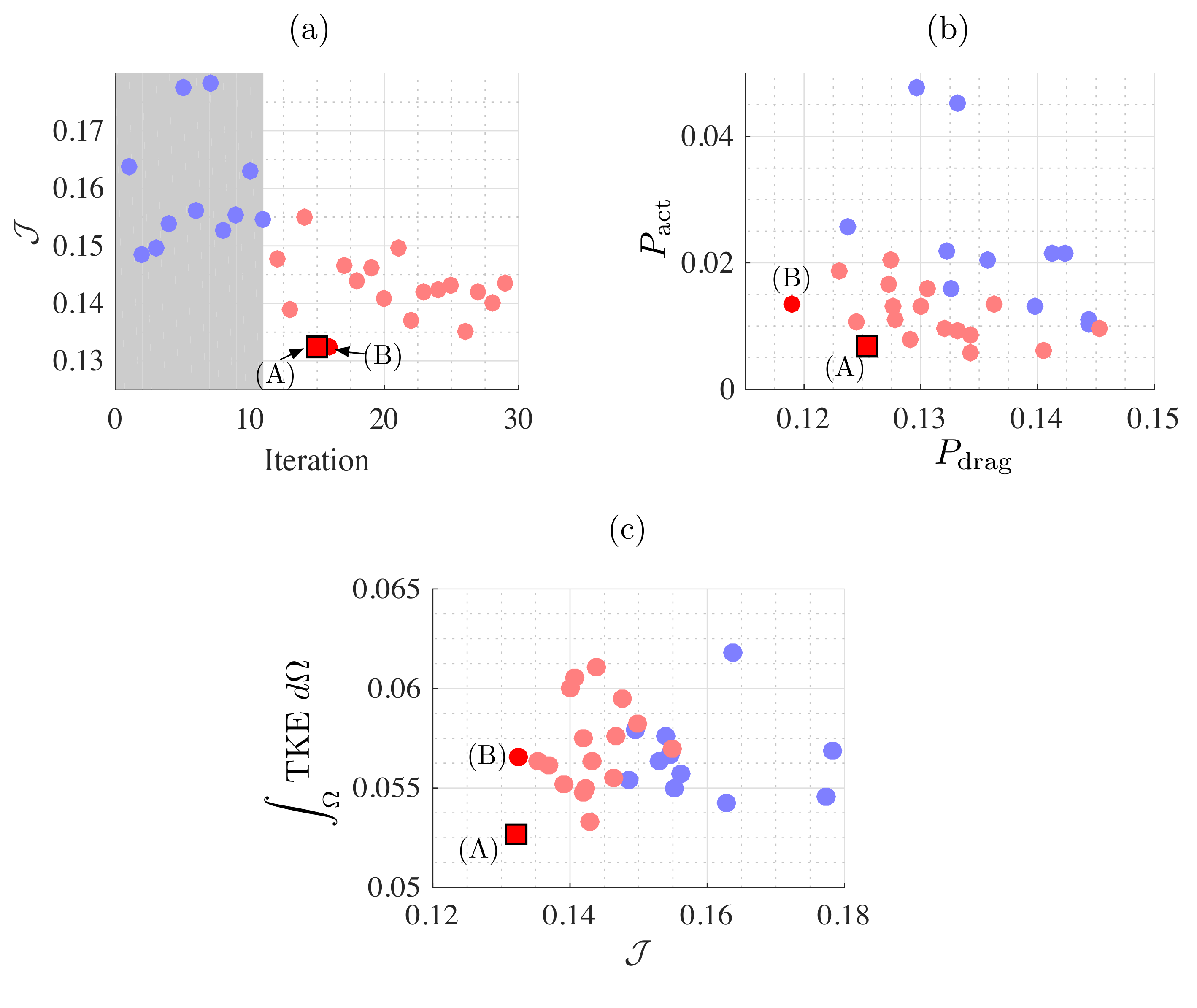} 
\end{center}
\caption{Control of 3D separated flow: (a) Objective function $\mathcal{J}$ minimization to determine the optimized control case, and (b) individual power consumptions $(P_\text{drag},P_\text{act})$. The square symbol denotes the optimized case. (c) Integrated turbulent kinetic energy (TKE) over the entire computational domain $\Omega$. The transparent blue dots indicate initial simplex control cases and the red square symbol denotes the optimized case. } 
\label{fig:3Dobjective}
\end{figure}

Time averages in the controlled flow are estimated over eight periods of dominant wake shedding frequency. The minimization of the cost function $\mathcal{J}$ associated with the control simulations is outlined in Figure \ref{fig:3Dobjective}(a). Following the initial simplex cases (shown in blue), the optimization procedure iteratively minimizes the total power consumption. The cost function associated with aerodynamic and actuation power is shown in Figure \ref{fig:3Dobjective}(b). Following the optimization procedure, the optimized control case A associated with minimum power consumption is deduced. We also highlight another control case B, whose cost function evaluation is similar to case A. Control case A is associated with lower actuation power ($P_\text{act}$) and correspondingly lower $\overline{C}_\mu = 0.0068$, while control case B is associated with higher $\overline{C}_\mu = 0.016$. However, the aerodynamic power ($P_\text{drag}$) is lower for B compared to A. In order to evaluate the effect of minimization of cost functional to the overall flow physics, we integrate the TKE in the computational domain $\Omega$ for all the control cases, which is shown in Figure \ref{fig:3Dobjective}(c). A minimum integrated TKE for the optimized control case A is obtained. Due to a lower actuation power input and lower velocity fluctuations in the streamwise wake associated with drag reduction, the TKE fluctuations are minimized with the optimized control law. 

\begin{figure}
\begin{center}
\includegraphics[width = 0.99\textwidth]{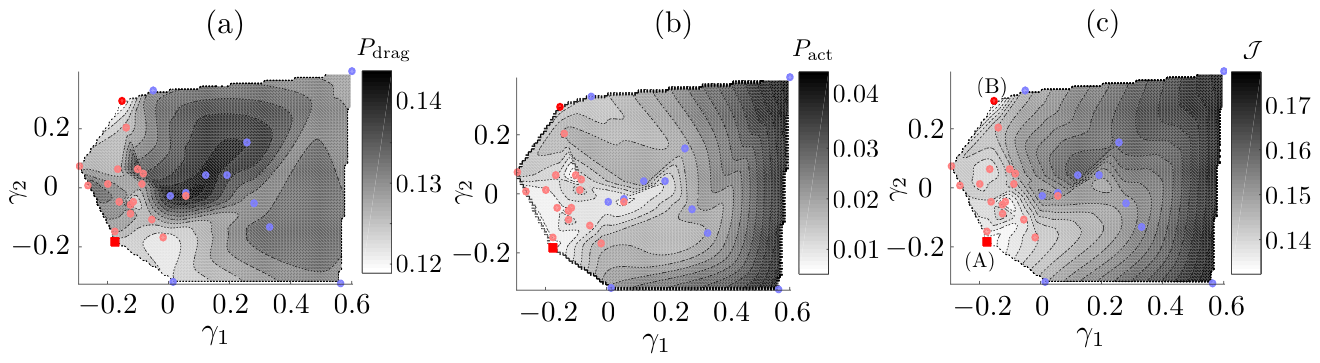} 
\end{center}
\caption{3D flow control landscape, (a) $P_\text{drag} = P_\text{drag}(\gamma_1,\gamma_2)$, (b) $P_\text{act} = P_\text{act}(\gamma_1,\gamma_2)$, and (c) $\mathcal{J} = \mathcal{J}(\gamma_1,\gamma_2)$, using multidimensional scaling. The transparent blue dots indicate initial simplex control cases and the red square symbol denotes the optimized case. } 
\label{fig:3Dobjective1}
\end{figure} 

The control landscape over the minimization variables is analyzed with MDS as shown in Figure \ref{fig:3Dobjective1}. Compared to the 2D control effort, the proximity map is more complex for 3D flow control. Although the actuation power $P_\text{act}$ increases for $\gamma_1>0$, the aerodynamic power $P_\text{drag}$ does not decrease correspondingly. The control landscape converges at the optimal location for the cost function $\mathcal{J}$ balancing both power considerations. Control cases A and B  occupy different positions in the proximity map shown in Figure \ref{fig:3Dobjective1}. The cluster-based procedure is able to extract not only the global minima but also the local minima, highlighted from different regions of the proximity map. For further optimization of the control laws with lower $\overline{C}_\mu$, a simplex consisting of $K+1$ cases near the current optimized case A can be chosen. For further optimization of the control laws with higher $\overline{C}_\mu$ and better aerodynamic performance, a simplex consisting of $K+1$ cases near case B can be chosen. The gradient-free searching algorithms for optimizing cluster control amplitudes explore different regions of the control landscape effectively. Thus, the proximity map can serve as a guide for tracking the performance of the flow control cases.  

\begin{figure}
\begin{center}
\includegraphics[width = 0.99\textwidth]{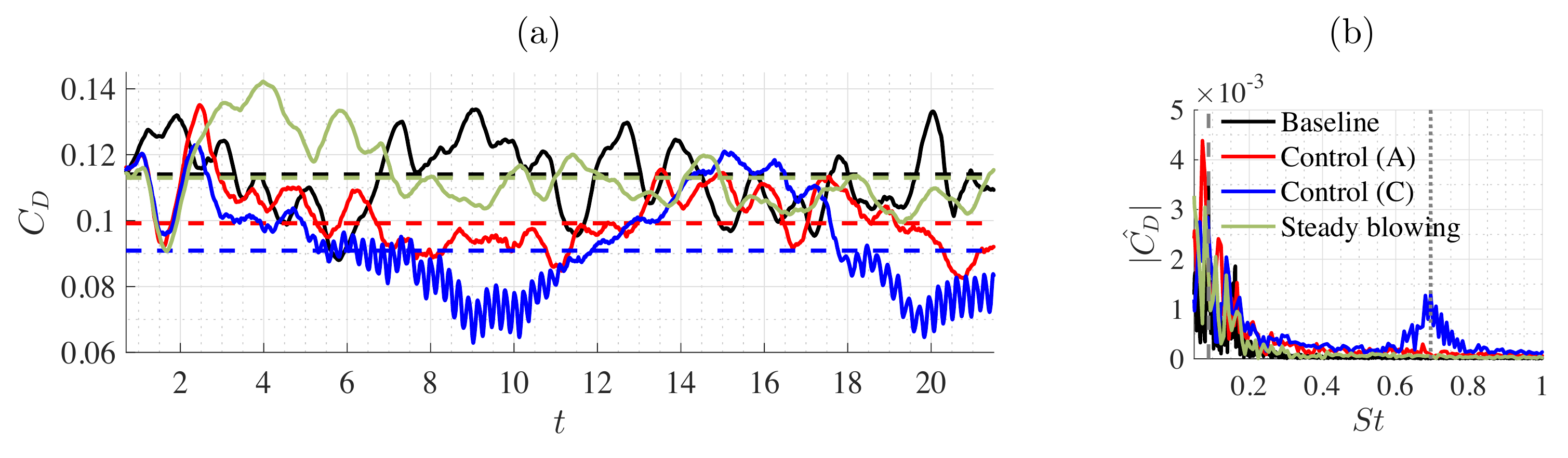} 
\end{center}
\caption{Comparison of baseline, the optimized 3D flow control case A ($\overline{C}_\mu = 0.0068$), the control case C with higher feedback gain ($ \beta = 1.6, \overline{C}_\mu = 0.016$) and steady blowing ($\overline{C}_\mu = 0.016$): (a) Drag coefficient, (b) spectral analysis of drag data. The dashed lines in (a) indicate the mean drag values. The dashed line in (b) corresponds to the dominant shedding frequency and the dotted line corresponds to the shear layer frequency.} 
\label{fig:3Dobjective2}
\end{figure} 

Using the same cluster control amplitudes $\{b_k^\text{opt}\}_{k=1}^K$ associated with the optimized case A, let us increase the feedback gain $\beta = 1.6$ in Eq.~\eqref{bhat} to evaluate the effect of increasing actuation input on control performance. This yields an increased $\overline{C}_\mu = 0.016$. This cluster-based control case with higher feedback gain will be referred to as case C in the following discussion. We also perform flow control with steady blowing at $\overline{C}_\mu = 0.016$ to compare with the cluster-based control cases A and C. The drag coefficient obtained in the three control cases is compared with the baseline as shown in Figure \ref{fig:3Dobjective2}(a). The black, red, blue and green dashed lines indicate the mean drag associated with the baseline, case A, case C and steady blowing, respectively. We obtain a $13\%$ drag reduction for case A and $20\%$ drag reduction for case C. We do not get any significant drag reduction with steady blowing. This is consistent with the work by \cite{munday2017effects}, where it was shown that the time-averaged separated flow was not significantly modified with steady blowing at $\overline{C}_\mu = 0.01$ and comparable drag reduction was achieved only at much higher $\overline{C}_\mu = 0.021$. Thus, the optimized cluster-based control laws, even with a lower $\overline{C}_\mu$ performs much better than steady blowing case. It must be emphasized that the objective of the control strategy is to minimize power consumption. With that objective, we are able to achieve drag reduction in 3D separated flows.

 A spectral analysis of the drag coefficient is performed to highlight the associated amplitude and frequency range of forcing, as shown in Figure \ref{fig:3Dobjective2}(b). Here, $|\hat{C}_D|$ is the single sided amplitude. For the optimized control case A,  the forcing is applied in the range of wake frequencies with the peak at $St^{+} = 0.0844$, close to the dominant shedding frequency. We notice that the single-sided amplitude near the wake frequencies for this case is much higher compared to the steady blowing case. The drag reduction obtained in case A is comparable with open-loop periodic forcing at the wake frequency examined in the work by \citet{amitay2002role}. Using a feedback control strategy yields a faster transient response for optimized control case A than the open-loop counterpart. For control case C, along with the wake frequencies, the dominant shear-layer frequency $St^{+} = 0.695$ is triggered with the cluster-based feedback control. The increase in feedback gain associated with the optimized control amplitudes results in enhanced aerodynamic characteristics, observed from the resulting drag reduction in Figure \ref{fig:3Dobjective2}(a). The fact that the cluster-based control strategy is able to  adaptively force the flow at characteristic frequencies corresponding to fundamental instabilities of the baseline flow purely from the feedback of select observables demonstrates the power of this data-based approach.

\begin{figure}
\begin{center}
\includegraphics[width = 0.95\textwidth]{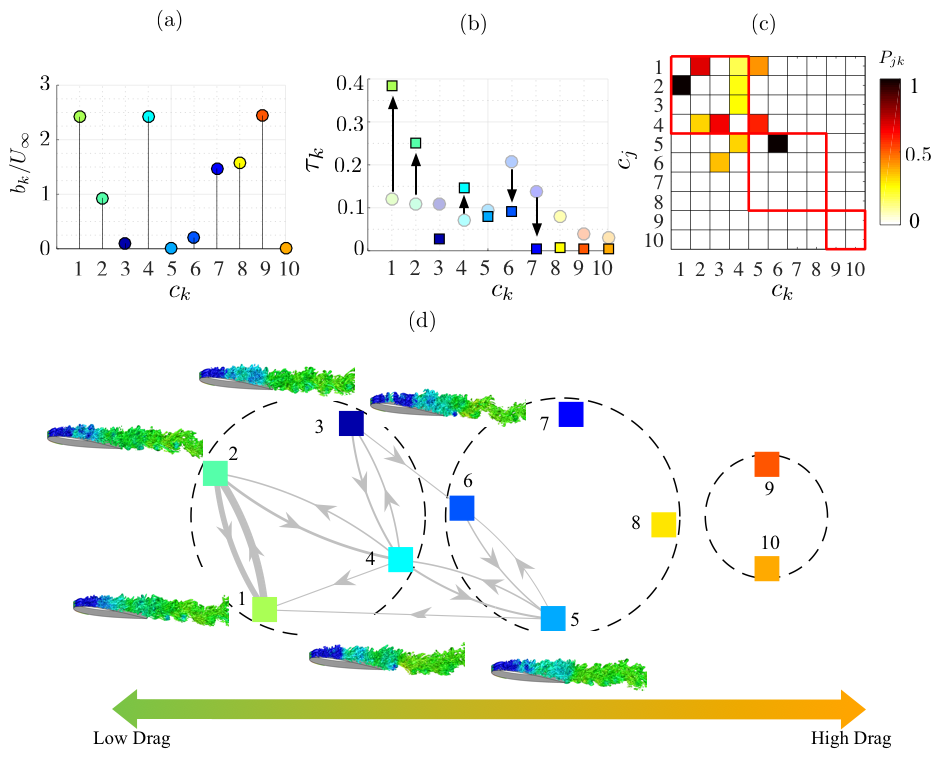} 
\end{center}
\caption{Comparison of optimized 3D flow control case A with baseline: (a) Cluster jet velocities ($b_k/U_\infty$), (b) cluster probability $\tau_k$ for baseline (shown in transparent round symbols) and controlled flows (shown in square symbols). (c) Cluster transitions for controlled case A, (d)  Markov chain associated with optimally controlled flow.}
\label{fig:3DMarkov}
\end{figure}

The cluster based control amplitudes $b_k$ corresponding to the optimized control case A is shown in Figure \ref{fig:3DMarkov}(a). Only positive amplitudes are present which indicate that only blowing is introduced by all clusters as constrained in the optimization procedure. Clusters $1$, $4$, and $9$ have maximum blowing amplitudes, followed by clusters $8$, $7$, and $2$. The remaining clusters $3, 5, 6$ and $10$ have considerably lower control amplitudes. The cluster probabilities for the controlled flow trajectories are compared with the baseline flow as shown in Figure \ref{fig:3DMarkov}(b). The controlled flow spends extended times in the low-drag states of clusters $1$, $2$ and $4$. The cluster probabilities associated with subset II clusters $6,7$ and $8$ reduce considerably. The cluster transitions in the controlled flow and the controlled Markov transition network are shown in Figure \ref{fig:3DMarkov}(c) and (d), respectively. Blowing in cluster $2$ deters transitions to cluster $3$. Also, blowing in cluster $7$ deters the transition $3\rightarrow7$. This eliminates the transition to the high-drag states of the flow leading to drag reduction and controlling flow separation. Natural baseline transitions are invoked in clusters $3$, $5$, and $6$ eventually redirecting the flow to low-drag cluster states. In addition to the above mentioned cluster transition dynamics, a cyclic transition between clusters $1$ and $2$ at shear layer frequency is obtained for case C.   

\begin{figure}
\begin{center}
\includegraphics[width = 0.99\textwidth]{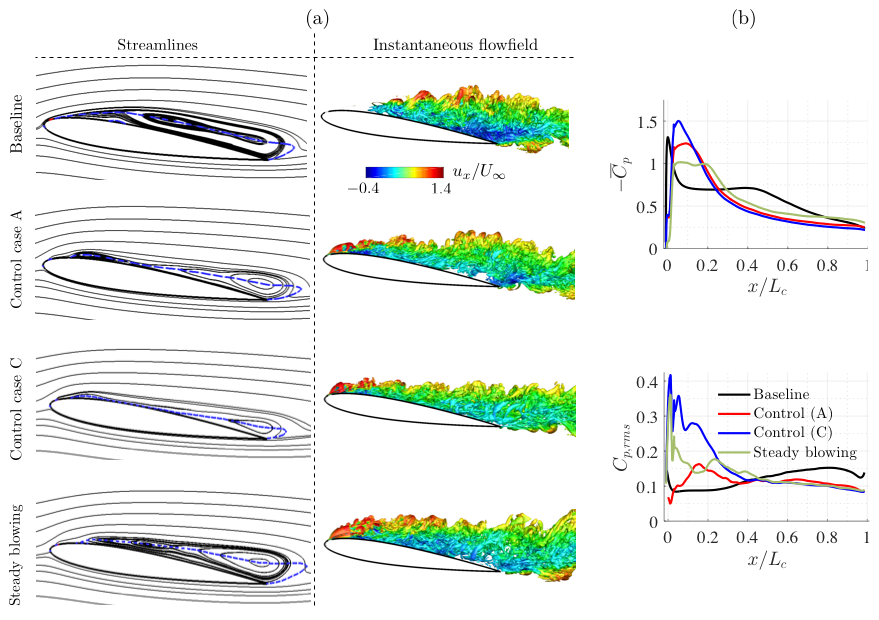} 
\end{center}
\caption{Comparison of baseline, optimized control case A ($\overline{C}_\mu = 0.0068$), control case C with higher feedback gain ($ \beta = 1.6, \overline{C}_\mu = 0.016$) and steady blowing at $\overline{C}_\mu = 0.016$: (a) time-averaged streamlines and instantaneous flow field (highlighted by $Q$-criterion) colored by streamwise velocity. The blue dashed lines indicate the contour line of zero time- and spanwise-averaged streamwise velocity. (b)  Time-averaged and root-mean-square fluctuation of coefficient of pressure distributions on suction side of the airfoil. }
\label{fig:3Dobjective3}
\end{figure}

We also show the streamlines and instantaneous flow fields associated with the baseline, control cases A, C and steady blowing case in Figure \ref{fig:3Dobjective3}(a). The blue dashed lines indicate the contour lines corresponding to time- and spanwise-averaged streamwise velocity, $\bar{u}_x/U_\infty= 0$. This characterizes the extent of the separation region in the flow. For both control cases, the size of the separation bubble is reduced compared to the baseline flow. For case A, the flow reattaches near the mid-chord of the airfoil. However, in this case, the flow separates near the trailing edge. Streamwise velocity deficit upstream translates to this trailing edge separation. In the control case C, the size of the separated region is considerably reduced. The vortical structures in the flow field are highlighted by a level set of the $Q$-criterion, which is colored by the streamwise velocity component. The vortical perturbations triggered by unsteady forcing at shear-layer frequencies result in a break-up of spanwise vortices. Suppression of separation due to a accelerated laminar-turbulent transition over the separation bubble is obtained, which results from the break-up of spanwise vortices. The coherence in the wake originating from von K\'{a}rm\'{a}n vortical structures is correspondingly lost leading to entrainment of free-stream momentum and mixing, which is consistent with the findings in \cite{Greenblatt:PAS00} and \cite{yeh2018resolvent}. The effectiveness of the cluster-based strategy can be highlighted by observing the time-averaged streamlines and instantaneous flow fields associated with the steady blowing case. Control case C reduces the separated region much more than the steady blowing case as it takes advantage of fundamental instabilities in the flow. It must be noted however that no a priori knowledge of such instabilities are provided for the flow control design.   

In both control cases A and C, we observe a low-frequency modulation in the drag coefficient. This modulation causes an increased unsteadiness in the drag force. These have been observed before in pulse-modulated actuated studies reported in \cite{amitay2006aerodynamic}. Such modulation results from a sub-optimal pressure recovery closer to the trailing edge. Previous studies with a model-based feedback control have shown that suppressing this low-frequency modulation can yield additional performance benefits \citep{nair2018networked}. There is an opportunity for a model-free extension to suppress the low-frequency modulation using the current feedback control strategy using additional clusters \citep{tesfaye2018clustering}. We also want to emphasize that complete flow reattachment is not the objective of the present work. The cluster-based control strategy is primarily designed to minimize aerodynamic power consumption. The optimized control laws yielded drag reduction and the reduction in the size of the separation bubble as additional benefits of the flow control strategy.       

The time average and root-mean-square fluctuations of the pressure coefficient distributions over the suction surface of the airfoil are shown in Figure \ref{fig:3Dobjective3}(b). For control case C, the average suction pressure is the highest at $x/L_c = 0.1$. For both control cases, the flat pressure region disappears indicating an accelerated laminar-turbulent transition. In contrast, for steady blowing, we see a flat pressure region indicating a slower laminar-turbulent transition. The accelerated roll-up and transition can be observed by peaks in the root-mean-square fluctuations of the pressure coefficient over the suction surface around $x/L_c = 0.15$.

With 3D flow control, we minimize the power consumption using the present cluster-based control strategy. As demonstrated, the optimized cluster-based control amplitudes lead to a transformed Markov transition network, essentially re-routing trajectories to desirable state-space regions. The model-free, iterative optimization of the global control law over the coarse-grained feature space leads to optimized cluster transitions and cluster probabilities that minimize the aerodynamic power consumption. The approach can easily be extended to achieve other desired performance objectives. Moreover, only the information of the feature space trajectories is required for both control and optimization of the feedback control laws.


\section{Concluding Remarks}
 \label{sec:cr}

We propose a feedback control strategy leveraging data-based clustering and optimization. The approach is demonstrated for two-dimensional (2D) and three-dimensional (3D) separated flows over a NACA 0012 airfoil at an angle of attack $\alpha = 9^\circ$, Reynolds number $Re = 23000$ and Mach number $M_\infty = 0.3$. The main objective of this study is to develop a model-free flow characterization technique and perform optimization of a global feedback control law, particularly to minimize the power consumption for aerodynamic flight.

The approach is based upon an unsupervised clustering of a feature space defined by the aerodynamic forces, $s^b(t) = (C_L(t),\dot{C}_L(t),C_D(t))$, collected from the baseline (unforced) large eddy simulations (LES). Centroid-based clustering analysis, performed using the $k$-means algorithm, segregates the characteristic phase regimes of the flow. These clusters partition the feature space into few discrete states. A statistical analysis of the transitions between these clusters yields linear Markov models, which characterize the coarse-grained, probabilistic dynamics of post-stalled flows. The different phases of vortex shedding can be analyzed by the characteristic flow fields associated with each cluster. Using a directed modularity maximization algorithm, groups of clusters (subsets) are extracted. These subsets divide the baseline trajectory into low, intermediate and high-drag states. For the Markov chain network associated with the 2D and 3D baseline flows, switching clusters are identified. These clusters are associated with key transitions from low to high-drag states. A modification of fundamental cluster transitions in a model-free manner is sought, which can be interpreted as a re-routing of trajectories associated with the baseline configuration to maximize desired performance objectives. 

Control amplitudes (blowing or suction jet velocity) are assigned to each cluster centroid for control. These cluster control amplitudes are interpolated over the entire low-dimensional feature space. A measurement of the current position on the trajectory relative to the cluster centroids is used to deduce a global control law. The control parameters in each cluster are then iteratively optimized to minimize power consumption. At each iteration step, the control law with the updated actuation parameters is evaluated in the simulation and the associated value of the cost function penalizing aerodynamic power and actuation power is determined. The optimization procedure yields a set of a control laws iteratively minimizing the cost functional. This optimization would be prohibitively expensive if performed on the full state-space; in contrast, our approach scales with the number of clusters. For 3D flow control, additional constraints were added to the optimization procedure to determine the unsteady feedback control laws for pure blowing. 

In the 2D flow control effort, a drag reduction of $41\%$ is achieved with the optimized feedback control law along with complete flow reattachment. The optimized cluster-based control law involves suction at the low-drag clusters and blowing at the intermediate and high-drag clusters. The optimized control law in 3D flow control achieves a $13\%$ drag reduction. This control law primarily operates in the range of wake frequencies associated with vortex shedding behavior. Drag reduction was accompanied by a decrease in the turbulent kinetic energy in the flow. Upon increasing the feedback gain associated with the optimized cluster amplitudes, vortical perturbations at forcing frequencies corresponding to the shear-layer instabilities are triggered in addition to the dominant wake frequencies. Although the actuation power increases with this feedback gain, an enhanced break-up of the spanwise vortical structures is obtained, yielding a $20\%$ drag reduction. Both cluster-based control cases perform significantly better than control with steady blowing. 

For both 2D and 3D optimized control cases, the cluster residence probabilities, i.e. the probability of the flow to remain in a particular cluster, of the low-drag states are increased with control and the probabilities associated with the high-drag states are decreased. The baseline Markov transition network is optimally modified with control to allow transitions that result in control of flow separation and drag reduction. It must be emphasized that the Markov models are only used as a tool for post-mortem simulation analysis making the current approach a model-free one. The optimization of the cluster-based control laws are typically achieved in limited runs, which scales as the number of discretized clusters.

In summary, the feedback control design using data-driven clustering provides a general, model-free and automated formulation for flow control. While we have demonstrated its applicability and power to extract optimized feedback control laws for separated flows, the present formulation is independent of the specific configuration and applicable to a variety of fluid flows with appropriate selection of feature space. Of crucial importance is the selection of the cost functional, as in any optimal control strategy. Moreover, limited number of sensor measurements are used for clustering and optimization which allows for the present method to be easily implemented in experiments. The cluster-based approach combines the modern day computing capabilities with data-driven techniques and can elevate future flow control efforts.       

%
\section*{Acknowledgements}

AGN, CY, SLB and KT are grateful for the support by the U.S. Air Force Office of Scientific Research (Award number: FA9550-16-1-0650; Program Manager: Dr. Douglas R.~Smith). AGN, CY and KT acknowledge the computational resources provided by the High Performance Computing Modernization Program at the U.S.~Department of Defense and the Research Computing Center at the Florida State University. AGN, BRN and KT also thank the international exchange program supported by U.S. Air Force Office of Scientific Research.  BRN acknowledges the support by the public grants overseen by the French National Research Agency (ANR) as part of the ``Investissement d’Avenir'' program, through the  ``iCODE Institute project'' funded by the IDEX Paris-Saclay (ANR-11-IDEX-0003-02) and ANR grants (`ACTIV\_ROAD' and `FlowCon').

\section*{Appendix A}
\label{form:ns2}

 To computationally examine the separated flows, (LES) are performed using a compressible flow solver CharLES \citep{bres2017unstructured}. The solver uses a second-order accurate finite-volume scheme and a third-order Runge-Kutta method for time integration. The computational domain is chosen to be $x/L_c \in [-19, 26],~y/L_c \in [-20, 20],~z/L_c \in [-0.1, 0.1]$,  following the work by \citet{yeh2017use}. To perform LES, Vremen's subgrid-scale model \citep{vreman2004eddy} is utilized. We apply Dirichlet boundary conditions of $[\rho,u_x,u_y,u_z,p] = [\rho_\infty,U_\infty, 0, 0, p_\infty]$ at the inflow and far-field boundary with a sponge zone specified near the outlet over $x/L_c \in [16, 26]$ to avoid numerical reflections \citep{freund1997proposed}. A no-slip adiabatic boundary condition is prescribed over the airfoil. A structured mesh consisting of $32$ million volume elements is utilized for the 3D LES simulations. The 2D LES simulations consider the same mesh discretization in the $x-y$ plane with $0.32$ million volume elements.  

The computational domain ($x-y$ plane) and mesh are shown in Figure \ref{fig:setup}(a) (left). The streamlines for the time-averaged 3D baseline separated flow is shown in Figure \ref{fig:setup}(a) (right). The chordwise direction is denoted by $\tilde{x}$. We notice the presence of a recirculation bubble in this separated flow. To characterize the extent of the separation region, a contour line of time- and spanwise-averaged streamwise velocity, $\bar{u}_x/U_\infty = 0$, is shown by the blue dashed line, which extends over the length of the airfoil. Here, $\bar{q}$ indicates time-average (mean) of flow variable $q$. The instantaneous flow field and turbulent kinetic energy (TKE) for 3D baseline flow are shown in Figure \ref{fig:setup}(b). TKE is defined as $\overline{({{u}_x^\prime}^2 + {{u}_y^\prime}^2 + {{u}_z^\prime}^2)}/U_\infty^2$, where $u^\prime \equiv u-\bar{u}$. The vortical structures in the flow field are highlighted by a level set of the $Q$-criterion \citep{hunt1988eddies}, colored by the streamwise velocity (${u}_x$). The laminar separation at the leading edge forms a shear layer that rolls up and evolves into spanwise vortical structures. TKE increases in this region of spanwise vortex formation reaching a maximum value at $x/L_c \approx 0.55$.

For validation of the numerical setup, we compare the time-averaged coefficient of pressure distribution ($\overline{C}_p$) for the 3D baseline separated flow with \cite{Kojima:JA13} in Figure \ref{fig:setup}(c), where agreement can be seen. In particular, a strong negative pressure peak on the suction side near the leading edge is visible followed by a plateau in pressure distribution. This pressure plateau indicates the presence of the separation bubble \citep{benton2018high}. For the 2D baseline simulation, the mean drag $\overline{C}_D = 0.127$ and the mean lift $\overline{C}_L = 0.818$. For the 3D baseline simulation, $\overline{C}_D = 0.114$ and $\overline{C}_L = 0.557$.

\bibliographystyle{jfm}
\bibliography{CBC_ref}

\end{document}